\documentclass[]{aa} 
\usepackage{graphicx}
\usepackage{txfonts}
\usepackage{xcolor}
\usepackage{comment}
\usepackage{hyperref} % added by James McK for links
\newcommand{\bl}[1]{\mbox{\boldmath$ #1 $}}

\begin{document} 

%\title{Formation of a dead zone and primordial dust ring in a gravitationally unstable protoplanetary disk}

   \title{Luminosity Outbursts in Interacting Protoplanetary Systems}

%  \title{The first generation of planetesimals in young gravitationally unstable protoplanetary disks}

%  \title{Formation of planetesimals in protoplanetary disks with a radially varying strength of gravitational instability }

%   \subtitle{VI. ???}

   \author{
    Aleksandr M. Skliarevskii\inst{1}, 
    Eduard I. Vorobyov\inst{2}
          }
   \institute{Research Institute of Physics, Southern Federal University, Rostov-on-Don 344090, Russia    
   \email{sklyarevskiy@sfedu.ru} 
   \and
	Ural Federal University, 19 Mira Str., 620002 Ekaterinburg, Russia
}
 
\date{12.02.2024}

%   \date{Received September 15, 2020; accepted November 16, 2020}
   
   \titlerunning{}
%   \authorrunning{Vorobyov et al.}

  \abstract
  {FU Orionis type objects (fuors) are characterized by rapid (tens to hundreds years) episodic outbursts, during which the luminosity increases by orders of magnitude. One of the possible causes of such events is a close encounter between stars and protoplanetary disks. Numerical simulations show that the fuor-like outburst ignition requires a very close encounter ranging from a few to a few tens of au. In contrast, the observed stellar objects in fuor binaries are usually hundreds of au apart. Simple mathematical estimates show that if such a close approach took place, the binary stellar components would have an unrealistic relative velocity, at least an order of magnitude greater than the observed velocity dispersion in young stellar clusters. Thus, the bursts are either triggered with a certain delay after passage of the periastron or their ignition does not necessary require a close encounter and hence the outburst is not caused by the primordial gravitational perturbation of the protoplanetary disk. In this work, an encounter of a star surrounded by a protoplanetary disk with a diskless external stellar object was modeled using numerical hydrodynamics simulations. We showed that even fly-bys with a relatively large periastron (at least 500 au) can result in fuor-like outbursts. Moreover, the delay between the periastron passage and the burst ignition can reach several kyr. It was shown for the first time by means of numerical modeling that the perturbation of the disk caused by the external object can trigger a cascade process, which includes the development of the thermal instability in the innermost disk followed by the magneto-rotational instability ignition. Because of the sequential development of these instabilities, the rapid increase in the accretion rate occurs, resulting in the luminosity increase by more than two orders of magnitude.}


   \maketitle

\section{Introduction}
It is known that protostars in young stellar systems can experience a sudden increase in luminosity. The luminosity during such events can change up to several orders of magnitude over timescales from tens to several hundred years. FU Orionis is the first object to successfully observe such an outburst. Subsequently, it formed the basis of a whole class of objects called fuors. Despite the fact that at the moment the number of objects classified as fuors reaches only a few dozen \citep[see, for example,][]{Audard2014, Magakian2022}, such outbreaks are hardly rare. Young stars can experience up to ten or even several dozen such events during their evolution \citep{Kenyon1999, Audard2014, VorobyovBasu2015}. Fuor-type outbursts are more likely to occur in the early stages of the evolution of protoplanetary systems, while there is still an accretion disk around the star, and this disk is active and optically thick \citep{Mercer2017, VorobyovBasu2015}.

Outbursts can affect the evolution of the disk, its structure and, in particular, its thermal characteristics. Even a short-term but significant increase in the luminosity of a star can noticeably heat the disk \citep{Vorobyov2014, VorobyovElbakyan2020}. In addition, the chemical composition of the disk can also be quite sensitive to both the heating of the disk and changes in the radiative characteristics of the star \citep{Visser2015, Rab2017, Molyarova2018, Wiebe2019, VorobyovBaraffe2013}. In addition to chemical reactions, temperature changes can result in a shift in the position of ice lines, a change in the properties of dust grains (for example, due to more active fragmentation) and, as a consequence, a change in the observed characteristics of the disk \citep{Banzatti2015, Schoonenberg2017}. For example, in \cite{Vorobyov2022}, it was shown that an outburst affects the distribution of spectral indices, and the effect persists for up to several thousand years. Finally, a sharp increase in the luminosity of the central star can serve as a trigger for various instabilities in the disk or change its dynamics. It was shown in \cite{Vorobyov2021} that the types of outbursts can be distinguished based on the dynamic characteristics of the post-outburst disk.

Despite the significance of outbursts for the evolution of the disk and their supposed prevalence, as well as the abundance of scientific works studying this phenomenon, it has still not been possible to reach a consensus in explaining the causes of outbursts. At the moment, many hypotheses have been presented about the physical nature of their origin. It is generally accepted that outbursts are a consequence of an episodic increase in the rate of accretion of matter from the disk onto the star \citep{Audard2014, Connelley2018}. Magnetorotational instability (MRI) is often considered as a phenomenon that potentially leads to a sufficient increase in the accretion rate \citep[see, for example,][]{Armitage2001, VorobyovKhaibrakhmanov2020}. This requires that the temperature in the disk be high enough to thermally ionize the alkali metals. In addition to the MRI, the accretion rate can increase when clumps of matter fall from the disk onto the star under the influence of gravity. Articles \cite{2010VorobyovBasu, VorobyovBasu2015} show that the accretion rate during such an event and the corresponding energy release during the fall of a massive clump onto a star can reach values characteristic of fuors. It is worth noting that not only clumps formed directly in the disk, but also compacted segments of the molecular cloud, for example, the remnants of the parent protostellar cloud, can fall on the star. The influence of the outer environment of the disk and accretion processes in the cloud-disk-star system was studied in \cite{Kuffmeier2018}. The capture of a large segment of the parent protostellar cloud by a protoplanetary disk was studied in \cite{Dullemond2019}. The fall of more compact remnants of protostellar clouds onto the central star was considered in \cite{Demidova2023}. The authors of these studies showed that, under certain conditions, such events can lead to the initiation of outbursts of the FU Orionis type. Mutual exchange of matter between the protostar and the low-orbiting planet may also cause a furore-like increase in the accretion rate and luminosity of the star, as proposed in \citep{Nayakshin2012}. In addition, thermal \citep{BellLin1994} and convective instability \citep[see, for example, ][]{Maksimova2020} claim to be the mechanism responsible for the increase in the accretion rate and the corresponding episodic increase in radiation. Finally, numerical simulations show that the passage of an outer star through a protoplanetary disk can cause a strong gravitational perturbation and cause an increase in the luminosity of stellar objects, similar to that observed for fuors \citep{BonnellBastien1992, Vorobyov2021, Dong2022}.

The FU Orionis object itself, which became the prototype of the fuors, is a double system consisting of the Northern and Southern sources. One of the stars, namely the less massive one, has been in an outburst phase since 1937, when its luminosity increased by two orders of magnitude on annual time scales. The second, more massive \citep{BeckAspin2012}, source remains in a quiet phase to this day. In light of this, it seems logical to assume that the origin of the FU Orionis type outburst is a consequence of the release of gravitational energy during the passage of the point of closest approach (periastron). However, close flybys in the plane of the disk or with small angles of inclination relative to the disk (which, apparently, is exactly what is observed in FU Orionis \citep{Perez2020}) require extremely close approach (from several AU to several tens of AU) and, accordingly, penetrating flight, which, according to the results of computer modeling \citep[see, for example, ][]{Vorobyov2021, Borchert2022, Cuello2023} is required to reproduce the outburst amplitude. When considering the spatial geometry of FU Orionis, it becomes clear that this assumption must be handled very carefully. This was noted, for example, in \cite{BeckAspin2012}. The problem is that the estimated distance between the FU Orionis stars is on the order of 250~AU \citep{Perez2020}. Following the logic presented in \cite{Liu2017}, we can estimate the average relative velocity of the sources, since we also know the time of the outbreak’s onset (i.e., the estimated moment of closest approach). Taking into account the above parameters, it can be calculated that the average relative velocity of objects should be at least 10 km/s, which is almost an order of magnitude higher than the velocity dispersion in young star clusters. This paradox can be resolved if we assume a large distance at periastron between the disturbing object and the central star (but in this case, it is problematic to restore the required outburst amplitude) or the presence of some delay between the closest approach and the initiation of the outburst. However, the factors that determine the length of the delay are still unknown. Moreover, the mechanism leading to the accretion burst in this case is likely to be different from direct gravitational interaction between objects. Thus, our goal is to study the possibility of the occurrence of outburst events in binary systems with large distances between the components in periastron (hereinafter referred to as major periastron) and delays in the initiation of bursts compared to the moment of periastron passage.

This paper presents the results of hydrodynamic modeling of a close approach of two subsolar mass objects with a high periastron. Following a non-penetrating flyby, an FU Orionis-type outburst occurs in the disk. The simulation shows a possible scenario for cascade initiation of an outburst, which may form the basis for explaining the nature of fuors in binary systems with large distances between components.

\section{MODEL OF THE CENTRAL PROTOPLANETARY DISK AND ITS PRE-OUTBURST STATE}
\label{sec:model}

To study outbursts in binary systems, numerical hydrodynamic modeling was used using the FEOSAD (Formation and Evolution of Stars and Disks) software code. This chapter will briefly describe the model used in the calculations, provide information about the initial conditions in the disk, and show the results of the simulation.

\subsection{Model Description}
The numerical hydrodynamic code FEOSAD, on the basis of which the work was carried out, makes it possible to simulate the coevolution of the disk and the central star in a two-dimensional thin disk approximation. Moreover, evolution can span hundreds of thousands and even millions of years. The main model used is presented in detail in \cite{2018VorobyovAkimkin}, as well as various modifications in articles \cite{Stoyanovskaya2020, Molyarova2021, VorobyovKhaibrakhmanov2020}. Next, the main components involved in the calculations and which are of particular importance for the current work will be briefly presented.

During the calculations, a system of hydrodynamic equations is solved that describe the joint evolution of the gas and dust components in the disk, namely:
\begin{equation}
\label{eq:cont}
\frac{{\partial \Sigma_{\rm g} }}{{\partial t}}   + \nabla_p  \cdot 
\left( \Sigma_{\rm g} {\bl v}_p \right) = 0,  
\end{equation}
\begin{equation}
\label{contDsmall}
\frac{{\partial \Sigma_{\rm d,sm} }}{{\partial t}}  + \nabla_p  \cdot 
\left( \Sigma_{\rm d,sm} {\bl v}_p \right) = - S(a_{\rm max}),  
\end{equation}
\begin{equation}
\label{contDlarge}
\frac{{\partial \Sigma_{\rm d,gr} }}
{{\partial t}}  + \nabla_p  \cdot 
\left( \Sigma_{\rm d,gr} {\bl u}_p \right) =  \nabla \cdot \left( D \Sigma_{\rm g} \nabla \left( {\Sigma_{\rm d,gr} \over \Sigma_{\rm g}} \right)  \right) +
S(a_{\rm max}),
\end{equation}

\begin{eqnarray}
\label{eq:mom}
\frac{\partial}{\partial t} \left( \Sigma_{\rm g} {\bl v}_p \right) +  [\nabla \cdot \left( \Sigma_{\rm
g} {\bl v}_p \otimes {\bl v}_p \right)]_p & =&   - \nabla_p {\cal P}  + \Sigma_{\rm g} \, {\bl g}_p + \nonumber
\\ 
&+& (\nabla \cdot \mathbf{\Pi})_p  - \Sigma_{\rm d,gr} {\bl f}_p,
\end{eqnarray}
\begin{eqnarray}
\label{eq:momDlarge}
\frac{\partial}{\partial t} \left( \Sigma_{\rm d,gr} {\bl u}_p \right) +  [\nabla \cdot \left( \Sigma_{\rm d,gr} {\bl u}_p \otimes {\bl u}_p \right)]_p  &=&   \Sigma_{\rm d,gr} \, {\bl g}_p + \nonumber \\
 + \Sigma_{\rm d,gr} \, {\bl f}_p + S(a_{\rm max}) {\bl v}_p,
\end{eqnarray}

\begin{equation}
\frac{\partial e}{\partial t} +\nabla_p \cdot \left( e {\bl v}_p \right) = -{\cal P} 
(\nabla_p \cdot {\bl v}_{p}) -\Lambda +\Gamma + 
\left(\nabla {\bl v}\right)_{pp^\prime}:\Pi_{pp^\prime}, 
\label{eq:energ}
\end{equation}
where are the components of the coordinate plane $(r, \phi)$ represented by indices $p$ and $p'$ . Equations \eqref{eq:cont} --- \eqref{contDlarge} are continuity equations written for gas, small (submicron) and grown dust grains, respectively. Wherein $\Sigma_{\rm g}$, $\Sigma_{\rm d,gr}$ and $\Sigma_{\rm d,sm}$ are the corresponding surface densities, and ${\bl v}_{p}$ and ${\bl u}_{p}$ --- velocity of gas and grown dust in the plane of the disk. Term $S(a_{\rm max})$ introduced to take into account possible transitions between fine and coarse dust populations. In turn, equations \eqref{eq:mom} and \eqref{eq:momDlarge} are equations for the dynamics of gas and grown dust. Note that there is no equation for the dynamics of fine dust, and in the continuity equation for fine dust the gas velocity is used, since the model assumes the joint movement of submicron dust with gas. $D$ — coefficient of turbulent diffusion, ${\bl g}_p$— components of the gravitational potential, including the contribution from the central star and the self-gravity of the disk, ${\bl f}_p$ —friction force between gas and dust, normalized per unit mass, $\mathbf{\Pi}$ — viscosity stress tensor, $\cal{P}$ — pressure integrated in the vertical direction. Finally, equation~\eqref{eq:energ} describes the energy balance of the system, where $\Lambda$ and $\Gamma$ are the rates of cooling and heating of the disk by radiation from the central star and background radiation, respectively.

The hydrodynamic model takes into account the following key processes: self-gravity of the disk (gas and dust); interaction between dust and gas through friction, including the reverse reaction of dust on gas \citep{Henderson1976, Stoyanovskaya2020}; turbulent viscosity, introduced according to the Shakura–Sunyaev parameterization \citep{1973ShakuraSunyaev}, and turbulent dust diffusion. The energy balance in the disk is calculated by taking into account energy changes due to compression and expansion (adiabatic heating/cooling), cooling by infrared radiation from the dust, and heating by radiation from the central star and background radiation. The dust component of the disk consists of two populations: fine dust, which is an ensemble of particles with sizes from 0.005~$\mu$m to 1~$\mu$m, and grown dust, whose minimum size is 1 $\mu$m, and the maximum is variable. Maximum size of grown dust amax changes due to the process of collisional growth through mutual collisions during turbulent and Brownian motion, as well as through advection. The size of dust is strictly limited by a fragmentation barrier, the magnitude of which depends on environmental parameters. Dust within populations is distributed according to degrees according to the law with an indicator $p = -3.5$, which corresponds to the infinite collisional cascade model \cite{Dohnanyi1969}. Transitions between populations are possible, with dust being redistributed so that the overall dust size distribution remains continuous at the point $a=1 \ \mu$m \citep[for more details, see ][]{Molyarova2021, Vorobyov2022}.

As mentioned above, the model involves $\alpha$-parameterization \cite{1973ShakuraSunyaev}. In this case, the viscosity parameter $\alpha_{\rm visc}$ is variable in space and time. The value $\alpha_{\rm visc}$ calculated in accordance with the “layered disk” concept  \cite{Gammie1996} using the approach \cite{Bae2014, Kadam2019}. In this case, the alpha value is determined as

\begin{equation}
\alpha_{\rm visc} = \dfrac{\Sigma_{\rm a} \alpha_{\rm a} + \Sigma_{\rm d} 
\alpha_{\rm d}}{\Sigma_{\rm g}},
\end{equation} 

where $\Sigma_{\rm a}$ and $\Sigma_{\rm d}$ — thicknesses of “MRI-active” and “MRI-dead” layers of the disk, respectively. Then $\alpha_{\rm a} = 10^{-2}$ is the value of the viscosity parameter in the active layer, and $\alpha_{\rm d}$ — in the remaining part of it, while $\alpha_{\rm d}$ also includes some background viscosity level
($\alpha_{\rm bg}=10^{-5}$). A description of the implementation of the method is given in more detail in \cite{Kadam2019}. Within the framework of the concept, it is assumed that regions with a depth of up to 100 g/cm$^{-2}$ are MRI-active, where the gas is sufficiently ionized by cosmic rays. It is also assumed that in regions of the disk in which the temperature exceeds the critical value $T_{\rm crit}$, thermal ionization of alkali metals occurs, which also leads to high MRI activity in these areas.

Finally, to simulate a close flyby of a (sub)stellar mass object, the approach presented in \cite{Vorobyov2017} is used, which allows us to simulate the collision of a protoplanetary disk surrounding the central star with an external source perturbing the system. Within the framework of this approach, an additional gravitational source is introduced into the system, which moves within the computational domain. The mass of the source can be large, especially in problems related to modeling the passage of external stellar objects. Consequently, the gravitational influence of the system on the central star cannot be neglected. Therefore, the simulation is performed in a non-inertial reference frame. Ultimately, the equation of motion in the system is modified by introducing additional components of the gravitational potential. To the initial components—the gravitational field of the star and the self-gravity of the disk—is added the gravity of the disturbing object and a term that allows us to directly take into account the movement of the reference system itself, by introducing an implicit potential.

The work will consider the results of two calculations: (i) the original model, which is a system with a single star surrounded by a disk; and (ii) a perturbed model in which an external object is launched towards the central star from a distance of 3000~AU. Radial and azimuthal components of the initial velocity of the external object $v_{\rm r}$ and $v_{\phi}$ are equal to $–2.5$ and $0.2$~km/s, respectively (a negative value here corresponds to movement towards the central star). The collision is retrograde, i.e., the disturbing object moves in the direction opposite to the rotation of the disk and the central star. The retrograde collision was chosen as more favorable for the development of FU Orionis-type outbursts at similar times scales, as shown in the work \cite{Borchert2022b}. The mass of the disturbing object is half the Sun, the mass of the disk surrounding the central star is 0.22~$M_{\odot}$, and the central star is 0.28~$M_{\odot}$. Thus, the mass of the outer object is higher than the mass of the central object on which the outburst is expected to develop, which is consistent with data on objects in the FU Orionis system \citep{BeckAspin2012}. The computational domain covers an area of $3500 \times 3500$~AU$^{2}$, divided into $400 \times 256$ cells in the radial and azimuthal directions, respectively. On the internal boundary of the settlement zone $r=0.2$~AU a boundary condition is established that allows the substance to freely flow in and out, while at the outer boundary, before the introduction of an external object, only the outflow of the substance is possible. However, when modeling a collision, the external boundary conditions also allow bidirectional movement of matter, i.e., both inflow and outflow. This is necessary to exclude the effects associated with the artificial accumulation of matter in the vicinity of the outer boundaries in calculations in a non-inertial reference frame \citep[for more details, see ][]{2017RegalyVorobyov}. 
 
\subsection{Initial State of Systems}
\begin{figure}
\includegraphics[width = \linewidth]{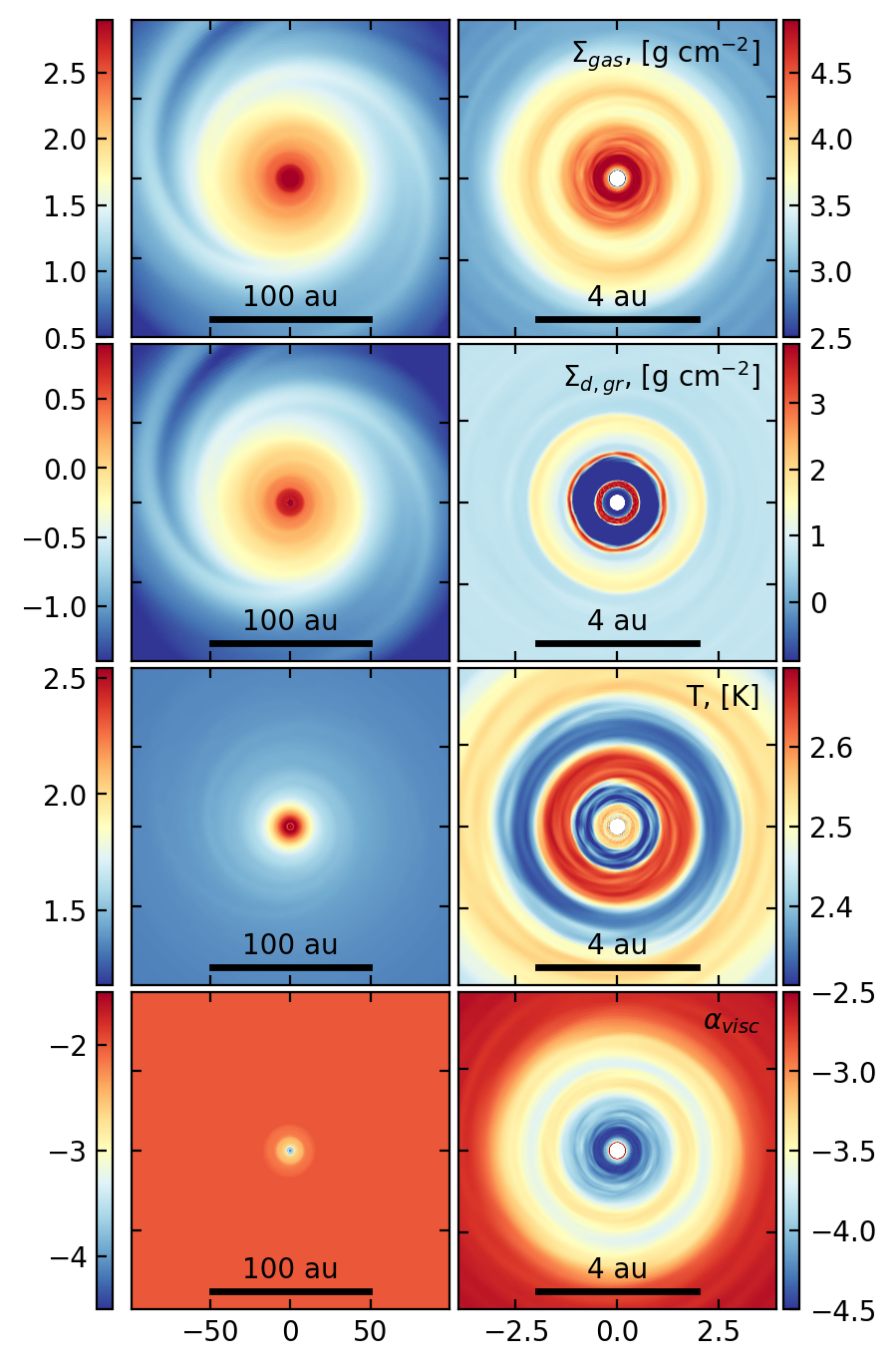}
\caption{Spatial distributions of the main parameters of the disk in the outer region $100\times100$~AU$^{2}$ (left) and internal area  $4 \times 4$~AU$^{2}$ (right). The values are presented from top to bottom: surface density of gas, surface density of grown dust, temperature, viscosity parameter $\alpha_{\rm visc}$. All values are shown in color on a logarithmic scale.}
\label{fig:initial}
\end{figure}

Protoplanetary disks, as well as the conditions in them at the start of calculations for both models, are equivalent. Initial distributions of gas surface density, temperature and viscous parameter $\alpha_{\rm visc}$ shown in Fig.~\ref{fig:initial}. These distributions were obtained by simulating the collapse of a prestellar cloud with a mass of $0.5 \ M_{\odot}$, disk formation and its subsequent evolution up to the point in time $t=125$ thousand years after
formation. First, let’s focus on the general structure of the disk, shown in the left column of Fig.~\ref{fig:initial}. It is clearly noticeable that the structure of the disk is non-axisymmetric, since it is gravitationally unstable. In this case, the mass of the outer shell remaining from the prestellar cloud exceeds 10\% of the total mass of the system. Consequently, the disk is in the embedded phase of evolution, for which the development of gravitational instability is quite expected \citep{2010VorobyovBasu}. The distribution of dust, in general, repeats the distribution of gas, although the surface density of the grown dust is approximately two orders of magnitude lower than the density of the gas. The large-scale thermal structure of the disk does not have significant features, and, despite slight deviations, is generally quite regular. According to the distribution of the viscosity parameter $\alpha_{\rm visc}$ shown in the bottom panel, the outer areas of the disk outside $\sim 15$~AU MRI-active. Within this region, a decrease in turbulent viscosity is observed, and fairly low values are achieved $\alpha_{\rm visc} \lesssim 10^{-3}$. 

Now let’s pay attention to the state of the internal disk, shown in the right column of Fig.~\ref{fig:initial} In internal 2 AU rings, both gas and dust, are clearly visible. In the case of gas, one of the rings is located at a distance $\sim2$ AU, and the second in close proximity to the internal boundary of the computational domain at a radius $\sim 0.35$ AU. In the dust component of the disk, 3 ring structures are observed: less dense at 2 AU, and more dense at 1 and 0.4 AU. At the same time, there is 1 a.u. inside. Between the outer and inner rings, as well as between the inner ring and the boundary of the computational domain, deep gaps are observed, in which the density of the grown dust drops by more than 2 orders of magnitude compared to the surrounding disk and rings.

Sharp transitions in the surface density of the gas are not random; their position coincides with transitions from higher to lower viscosity when moving from the outer part of the disk to the inner one, and the maximum gas density, accordingly, falls on the minimum $\alpha_{\rm visc}$. This can be seen in the bottom panel of Fig.~\ref{fig:initial} which shows the distribution of the viscosity parameter $\alpha_{\rm visc}$. Viscosity serves as one of the main agents of matter transfer in the disk. Therefore, when a substance from a zone with a high transfer rate (i.e., with a high viscosity parameter) enters a region with a low transfer rate of substance, the viscosity cannot ensure complete and unimpeded transport, which inevitably results in the accumulation of gas at the boundary between these two regions. In areas where gas accumulates at elevated temperatures, pressure peaks form, towards which dust drifts, thus collecting into the ring structures described above. The mechanisms of accumulation of matter and the formation of gas and dust rings are described in more detail in ~\cite{Kadam2019, Kadam2020, Vorobyov2023}.
 
 \section{FLYING BY A MASSIVE OBJECT AND ITS EFFECT ON THE EVOLUTION OF THE DISK}
 The state of the protoplanetary disk described in the previous chapter is used as initial conditions in two model cases. In the first case, the evolution of the system occurs without disturbances, and this model is accepted as a reference. In the second case, an external disturbing object with a mass is introduced into the presented system $M_{\rm int} = 0.5$~$M_{\odot}$, which is launched at a distance of 3000 AU. from the central star with radial and azimuthal velocities $v_{\rm r} = -2.5$~km/s and $v_{\rm \phi} = 0.2$~km/s. Note that, due to the two-dimensional approximation used, the passage occurs in the plane of the disk.
 
\subsection{Flash due to Primary Gravitational Disturbance}
\begin{figure}
\includegraphics[width = \linewidth]{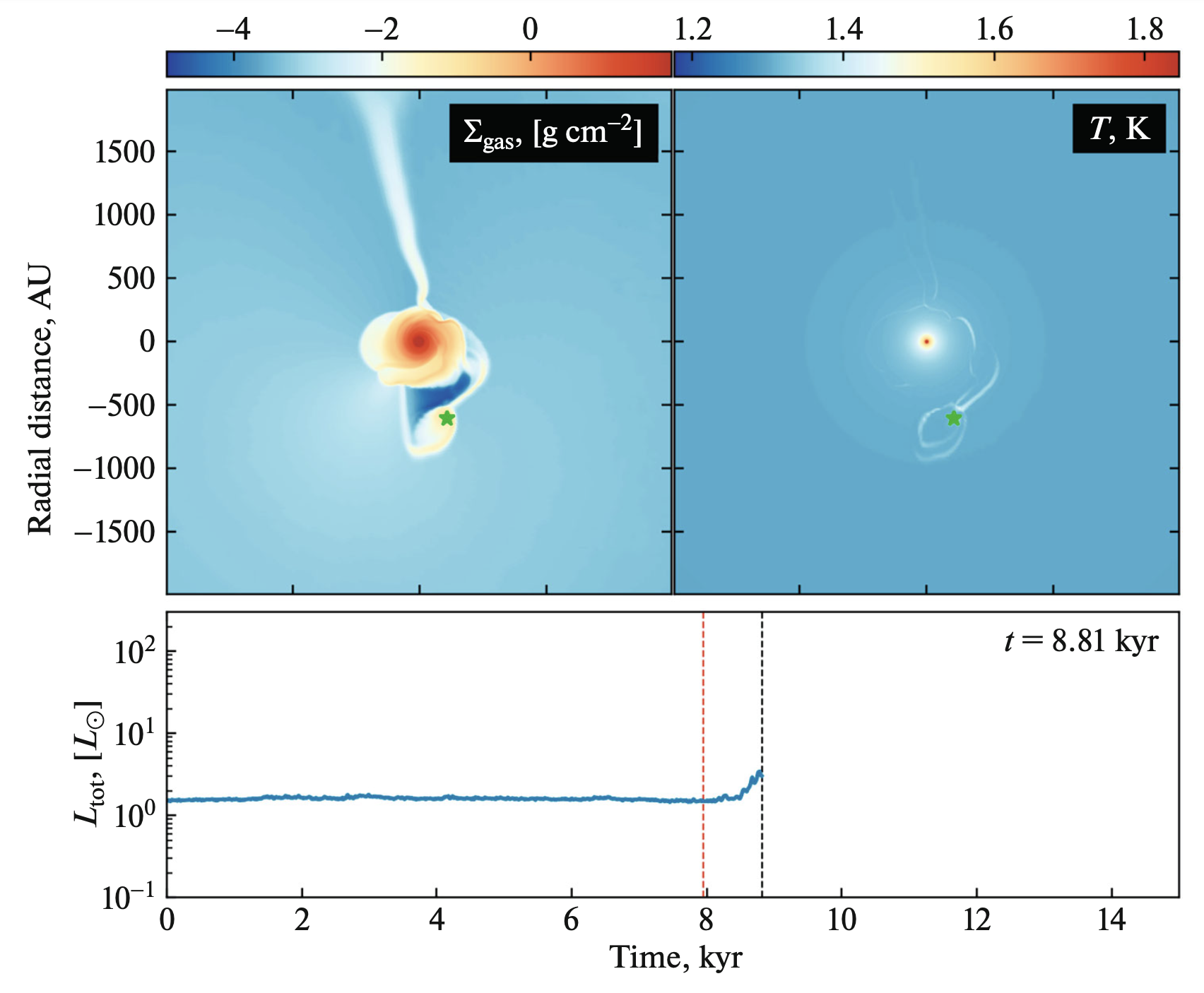}
\caption{Top panel: spatial distributions of gas surface density and temperature in the model with the passage of an external object, shown in the area $2000 \times 2000$~AU$^{2}$ at the moment the luminosity of the central star increases due to gravitational disturbance caused by the flyby. The bottom panel is a graph of the luminosity of the central star versus time from the start of the calculation to the moment shown in the top row of panels. The black vertical dotted line indicates the time t = 8.81 thousand years, corresponding to the local peak of luminosity, and the red dotted line is the moment of passing the point of closest approach—periastron.}
\label{fig:periastron}
\end{figure}
 The external object takes 7.95 thousand years to reach the periastron point. During this time, the external object manages, through gravity, to accumulate matter around itself from the remnants of the prestellar cloud, forming some kind of disk, despite the fact that full accretion onto the disturbing object is not included in the model. The minimum distance at which objects approach each other is approximately 500 AU. At the same time, at the moment of periastron passage there are no changes in the luminosity of the star, since the disturbance in the disk caused by the gravitational influence of the external object takes time to reach the accretion region. An increase in the luminosity of the central object, provoked by a change in the accretion rate caused by a disturbance from an external object, occurs over time $t=8.81$~kyr from the moment of its launch. Thus, the time required for the primary disturbance to travel through the entire space of the disk is about 860 years. The luminosity of the central object increases by 2–2.5 times. Although this change leads to a local increase in temperature in the inner disk (by 100–200 K in the innermost disk), the object’s luminosity remains far from that characteristic of FU Orionis objects, equal to 30 -- 300~$L_{\odot}$, \citep{Connelley2018}. Thus, the primary disturbance arising from the gravitational influence of an external object during the passage of the periapsis is not enough to increase the accretion rate similar to the fuor outburst in the presented case. 
  
\subsection{Secondary Outbreak}
\begin{figure*}
\includegraphics[width = \textwidth]{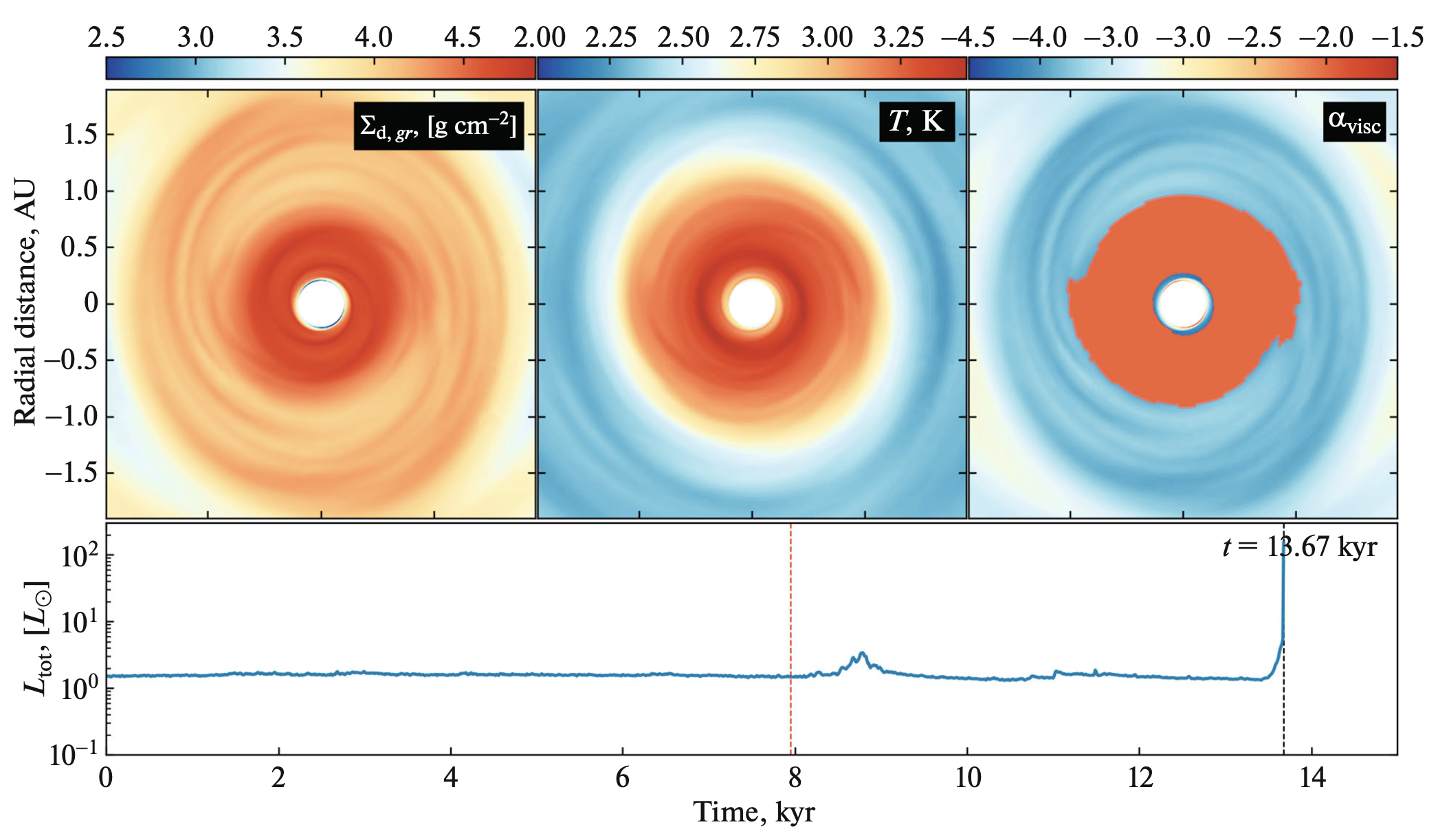}
\caption{Same as in Fig.~\ref{fig:periastron}, but during an outbreak of MRI, in the inner region $2 \times 2$~AU. In addition, the upper right panel also shows the spatial distribution of the viscosity parameter $\alpha_{\rm visc}$}
\label{fig:varphi}
\end{figure*}

\begin{figure}
\includegraphics[width = \linewidth]{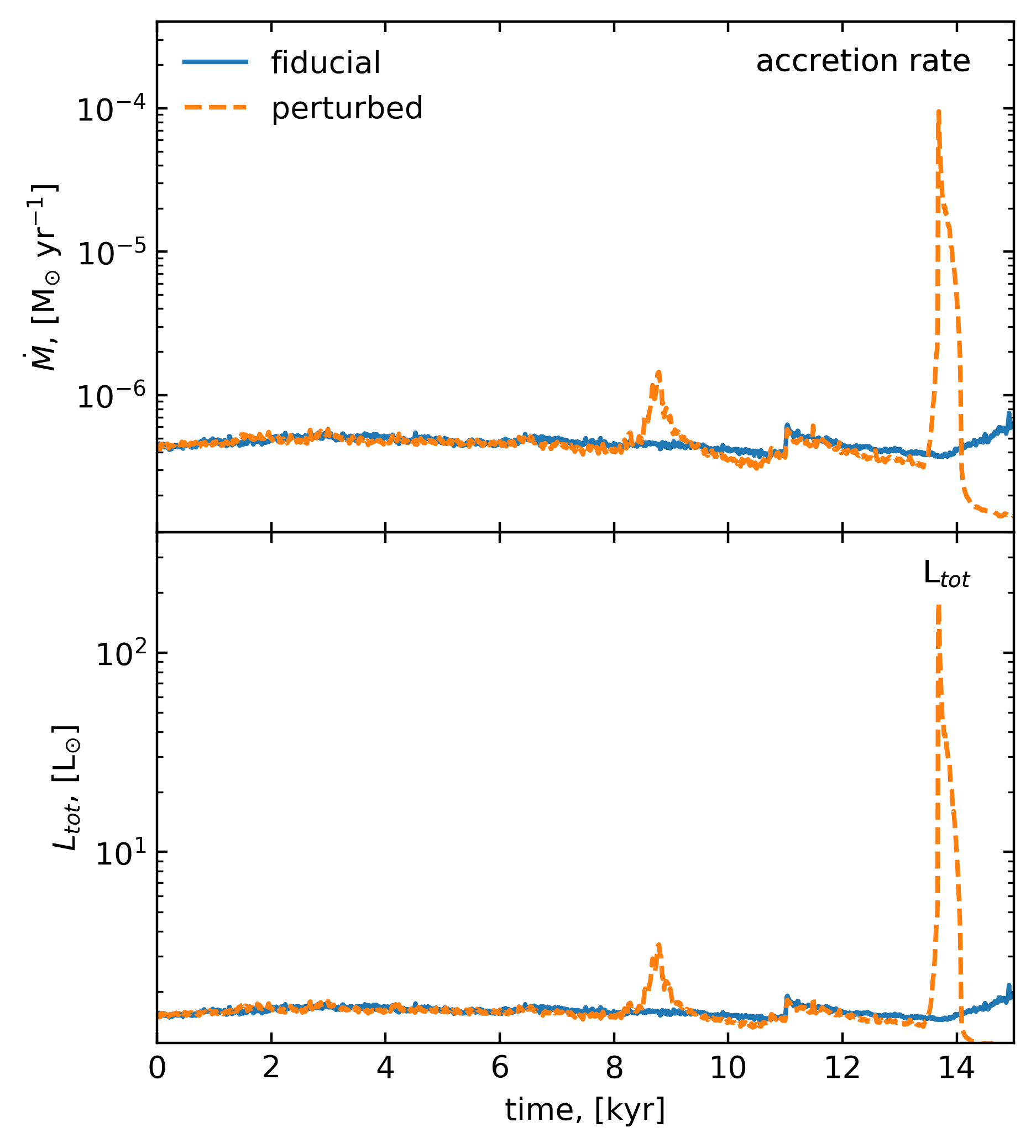}
\caption{Top: Rate of accretion of material from the disk onto the star as a function of time. Bottom: The total luminosity (as a combination of accretion and photospheric luminosity) of the central star. The blue solid lines correspond to the reference model, and the orange dashed lines correspond to the perturbed model.}
\label{fig:arates}
\end{figure}
However, the FU Orionis outburst still occurs 5.7 thousand years after the passage of periapsis. In this case, the cause of the outburst is not the gravitational effect, but the development of MRI in the inner disk. Figure~\ref{fig:varphi} shows the distribution of parameters in the internal disk. A sharp increase in temperature above the critical value and the viscosity parameter to the value $\alpha_{\rm visc} = 10^{-2}$ indicates the activity of MRI, which leads to a significant increase in the rate of matter transfer in the disk. This, in turn, results in a high rate of accretion of matter onto the star and, as a consequence, high luminosity of the central star. During this event, the luminosity rises by 2 orders of magnitude in a short period of time, less than 10 years, which is consistent with the characteristics of fuors. The frequency of data output in model calculations, however, does not make it possible to more accurately determine the time of peak luminosity. A fleeting increase in luminosity is followed by a gradual decay. The luminosity returns to pre-outburst values within about 400 years. The decay characteristics also correspond to current concepts and data on fuors. In particular, for the parent object of the class—FU Orionis— the rate of luminosity decay is approximately 2.5–3 times per 80 years \citep[see, for example, ][]{Labdon2021, Lykou2022}. In the presented model, the luminosity decreases by a factor of 2.65 over the same time period.
 
It is also worth noting that in the reference model no outbursts occur, as can be seen in Fig.~\ref{fig:arates}. This allows us to conclude that it is the passage of an external object that triggers the mechanisms that lead to the initiation of an outburst.

\section{DEVELOPMENT OF MAGNETOROTATIVE INSTABILITY DUE TO NON-PENETRATING FLYING} 

\begin{figure}
\includegraphics[width = \linewidth]{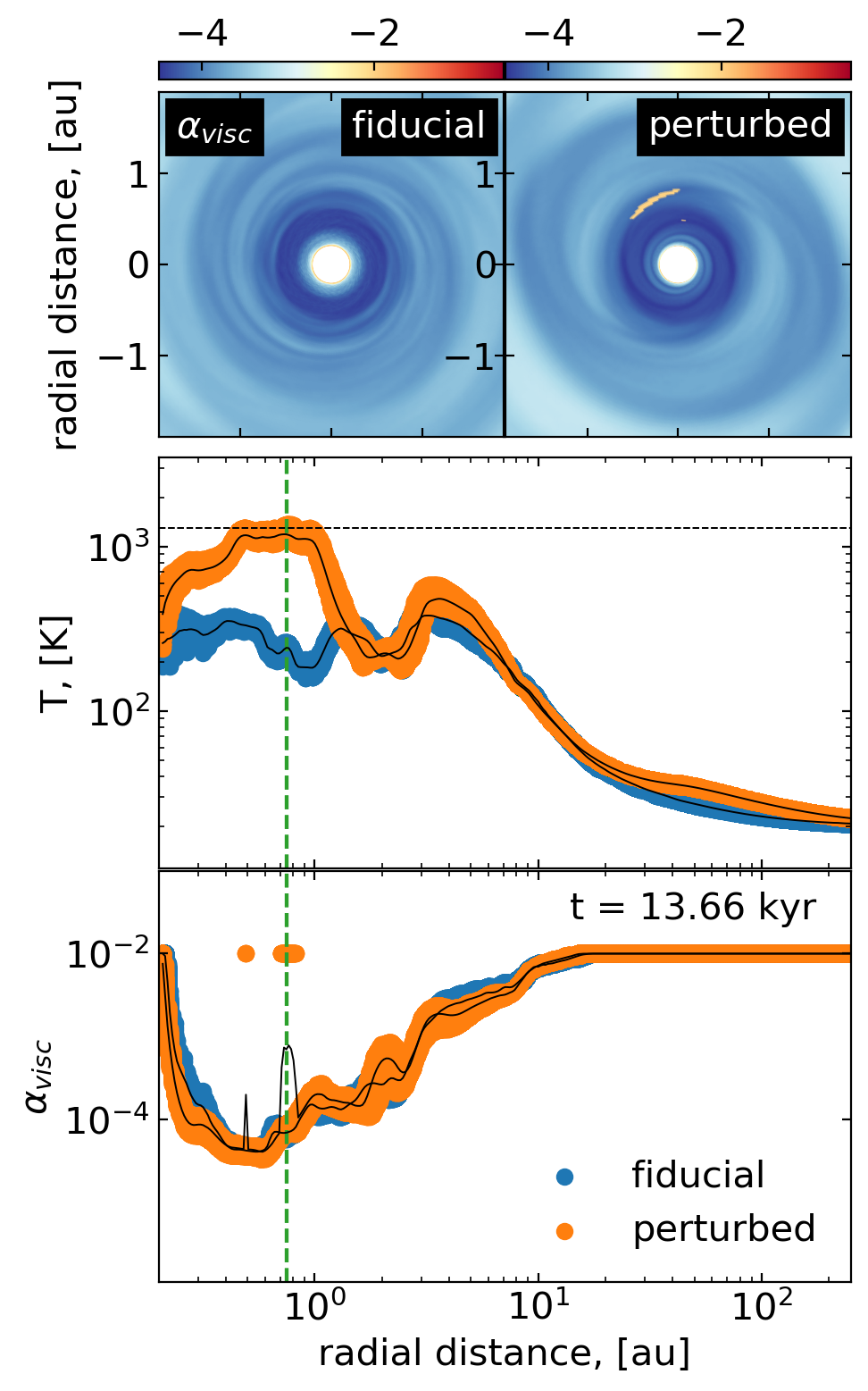}
\caption{State of disks in models at a point in time $t=13.66$ thousand years. Top panel: two-dimensional distribution of the viscosity parameter of the reference (left) and perturbed (right) models in the internal region $2 \times 2$~au. The temperature distribution profiles are presented in the middle panel, and the viscosity parameter profiles are presented in the bottom panel. The set of dots corresponds to all azimuthal values at a given radius, while the solid lines show the average values. Blue and orange points correspond to the reference and perturbed models. The green vertical dashed line marks the radial distance of 0.75~AU}
\label{fig:pre-burst}
\end{figure}

Let us consider the state of the disk immediately preceding the outburst of luminosity. Figure~\ref{fig:pre-burst} shows two-dimensional distributions of the viscosity parameter $\alpha_{\rm visc}$ in the inner region 2~AU. These distributions for the original and perturbed models are shown in the top row of panels on the left and right, respectively. Selected time $t = 13.66$ thousand years from the beginning of the calculation, i.e., 100 years before the outburst of luminosity. Dead zones characterized by low $\alpha_{\rm visc}$, are present in both cases, but in the perturbed model the dead zone is less symmetrical. Outside the narrow ring, in which the viscosity parameter is minimal, the distribution morphology also differs. While in the original model the morphology has a complex spiral nature, in the perturbed model a pattern in the form of an oblate ellipse can be traced. What is most important: in the considered area there are zones with a high value $\alpha_{\rm visc}$, corresponding to the developed MRI. It is worth noting that MRI in the disk develops somewhat earlier than the luminosity burst, but it should be clarified what leads to its development.

According to the layered disk model used, MRI develops in the outer disk as a result of the medium achieving a sufficient degree of ionization through ionization by cosmic rays. However, as mentioned above, the inner region is characterized by a high surface density, which means that only the uppermost layers of the disk are ionized, constituting only a small part of its thickness. On the other hand, the required degree of ionization can be achieved through the mechanism of thermal ionization. This requires that the temperature in the disk exceed a threshold value (in this study it is set equal to 1600 K \citep{Bae2014}). One-dimensional distributions of temperature and viscosity parameter are demonstrated in the lower panels of Figure~\ref{fig:pre-burst}. Along the vertical axis, colored dots show the sets of all values of quantities at a given radius. It is obvious that in some areas the temperature exceeds the critical value shown by the horizontal dashed line. It is in these regions that MRI develops. Thus, a luminosity burst develops in the perturbed model due to the activation of the MRI in regions of increased temperature in the disk. However, the question remains open as to what exactly causes the temperature to rise.

\subsection{Thermal Instability}
\begin{figure}
\includegraphics[width = \linewidth]{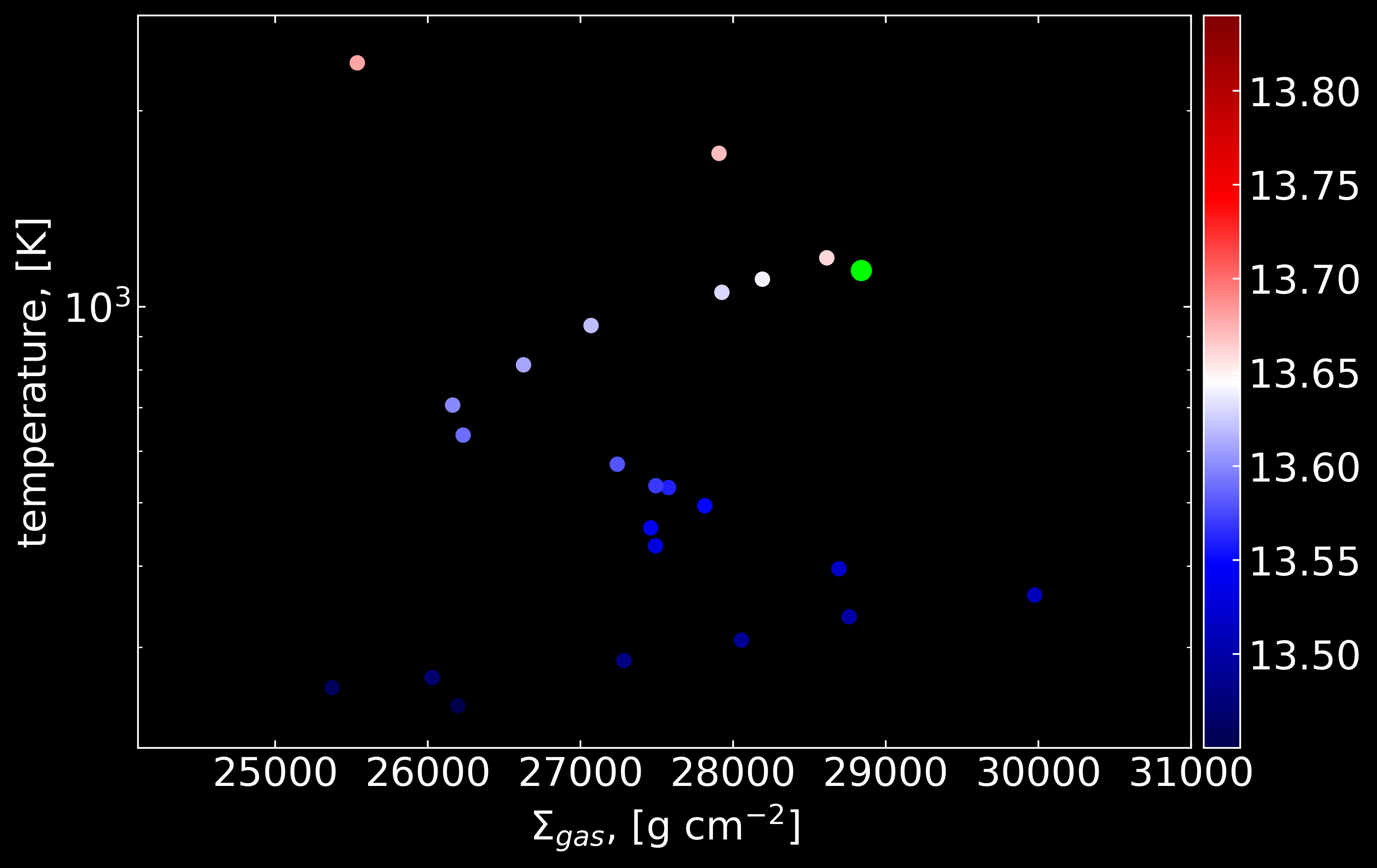}
\caption{Evolution of the state of a perturbed disk in the “surface gas density–temperature” phase space. The dots show states at different points in time. Time difference between the points is 10 years. Absolute time is shown in different colors. Shades of blue are for the times before the outbreak, red—after it. The green dot marks the moment immediately preceding the flash.}
\label{fig:s-curve}
\end{figure}

\begin{figure}
\includegraphics[width = \linewidth]{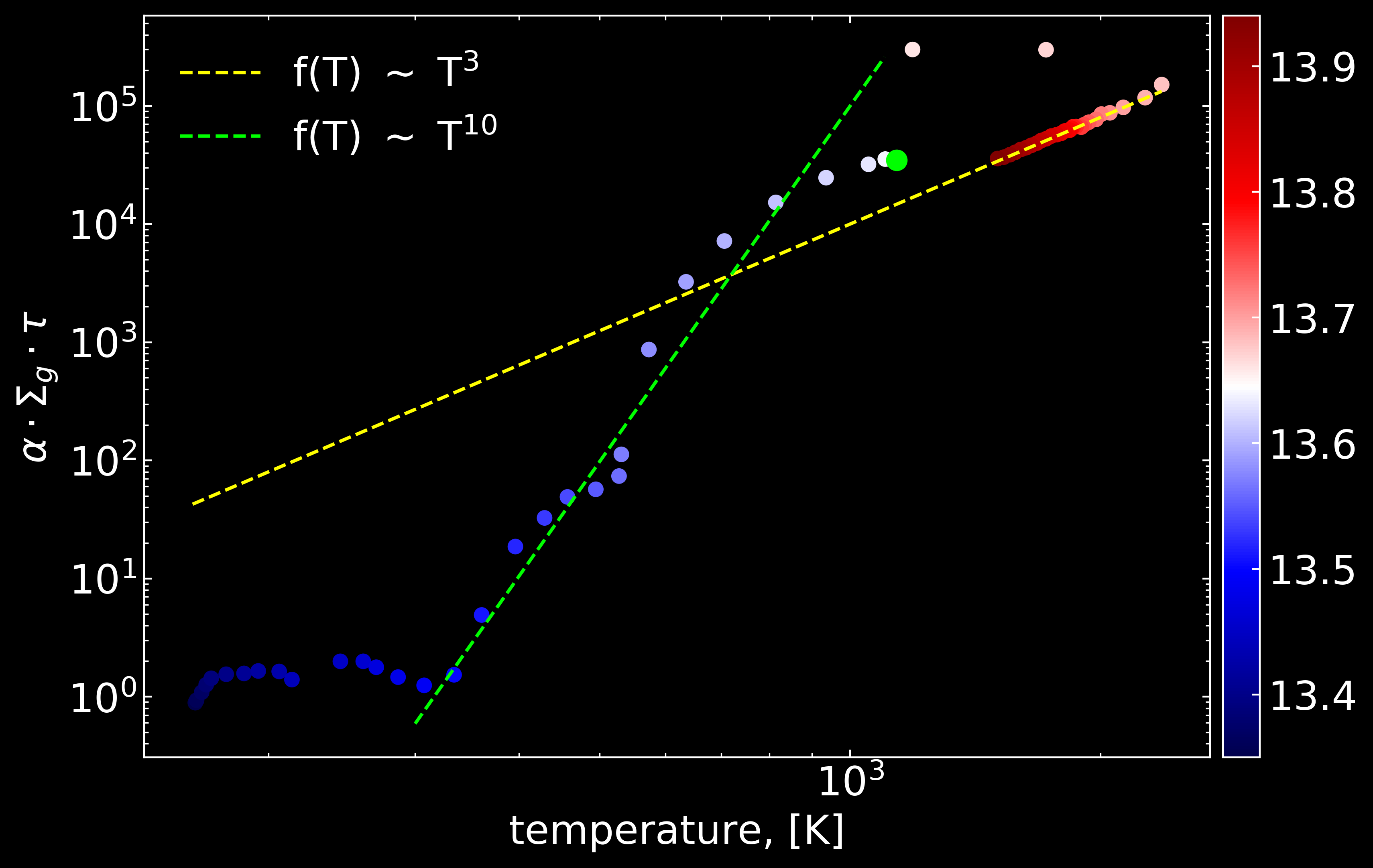}
\caption{Same as Fig.~\ref{fig:s-curve}, but in phase space “$T-\alpha_{\rm visc} \Sigma_{\rm g} \tau$}”.
\label{fig:heatrate}
\end{figure}

\begin{figure}
\includegraphics[width = \linewidth]{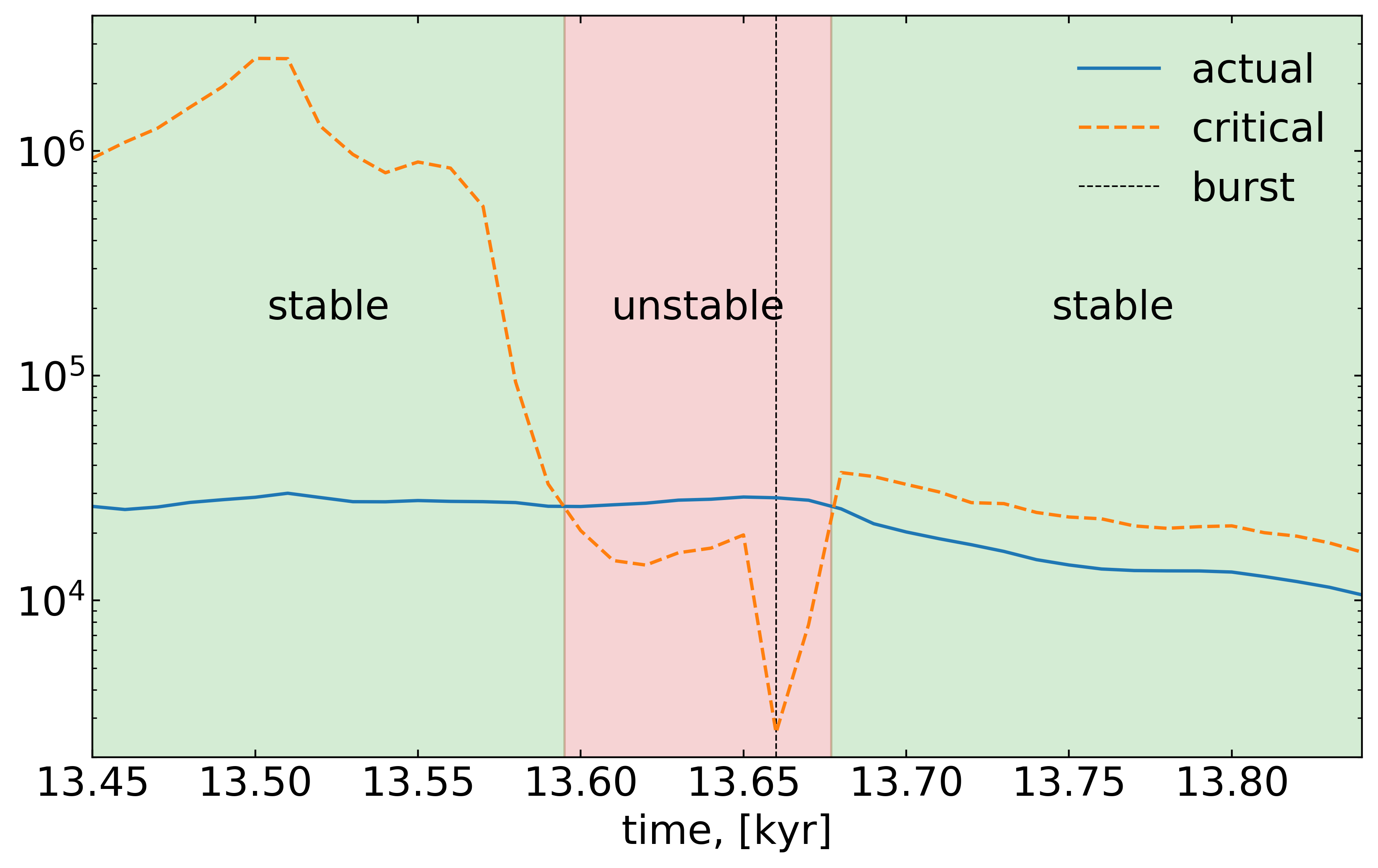}
\caption{Evolution of surface gas density values in a disturbed disk at times close to the outburst of luminosity. The blue solid line shows the actual values of the gas surface density. The orange dashed line indicates the critical value of the surface density, above which thermal instability develops in the disk. The vertical black dashed line shows the time point corresponding to the initiation of the flash.}
\label{fig:critdens}
\end{figure}
Let’s focus on the region of the disk located at a distance of 0.75 AU from the star, since at this distance the MRI develops first (the corresponding region is marked in the lower panels of Fig.~\ref{fig:pre-burst} with a green dashed line) and we will consider in detail the evolution of various parameters in this region immediately before, during and after the outburst. Since we are primarily interested in the thermal and accretion characteristics of the medium, it will be useful to trace the mutual changes in the surface density and temperature of the gas. Figure\ref{fig:s-curve} shows the evolution of the state of the selected region in the phase space “$\Sigma_{\rm g} - T$”. Time is shown using a color scale covering some period before (200 years) and after the outbreak (30 years). Each point represents the azimuthally averaged characteristics for a selected point in time. Time step between points is 10 years. In this case, the pre-outburst state corresponds to points of blue color and its shades, and the post-outburst state - red. The moment immediately preceding the f lash is marked with a green dot. In the considered time interval, the temperature continuously increases until the threshold value of MRI activation is reached $T_{\rm crit} = 1600$~K. In this case, the evolution of the surface density of the gas has a non-monotonic character, that is, the values either increase or decrease. In fact, for the state immediately preceding the outbreak, the curve shown has a curved shape, reminiscent of the Latin letter S. This shape of the curve in the chosen phase space is characteristic of systems subject to the development of thermal instability, described, for example, in \cite{BellLin1994, Hartmann1998, Kadam2020}. Let’s take a closer look at this feature.

In the works mentioned above, the thermal equilibrium curves have the shape of the letter S. That is, curves describing the state in which the main sources of accumulation and loss of thermal energy are viscous heating, dominant in the considered internal, optically thick region \citep[see, for example, ][]{Skliarevskii2021}, and cooling due to thermal radiation of dust from the surface of the disk, balance each other.

In the region to the left of the S-curve, cooling dominates over heating, and if the temperature rises, regardless of the cause, the system self-stabilizes and returns to an equal state. At the same time, heating dominates in the region on the right, which means that an increase in temperature results in subsequent additional growth, which continues until a new equilibrium state is reached. This process is called thermal instability.

Let us recall that Figure~\ref{fig:s-curve} shows the evolution of azimuthally averaged values. Based on the characteristic shape of the distribution, we can assume that in general the system evolves in accordance with the equilibrium curve. However, it is worth remembering that local temperature changes in the pre-outburst period, associated, for example, with local adiabatic compression caused by hydrodynamic f lows, provoke a transition to a new equilibrium state with an even higher temperature.

In a simplified form, the condition describing the equilibrium state of the system can be written as \citep{Kawazoe1993, DAlessio1996}:
\begin{equation}
\alpha_{\rm visc} \cdot \Sigma_{\rm g} \cdot \tau \propto T^{3},
\end{equation}
where  $\tau = \kappa \Sigma$ is the local optical thickness in the disk in the vertical direction, which is essentially the product of the opacity of the substance in the medium $\kappa$ and surface density of this substance $\Sigma$. Moreover, both gas and dust can act as the dominant carrier of opacity, depending on the temperature regime.

This expression is obtained by balancing the cooling and heating rates. If the product of parameters on the left side increases faster than the cube of temperature, this creates conditions for the development of thermal instability. This situation is exactly realized in the region under consideration, which can be clearly seen in Fig.~~\ref{fig:heatrate}. The figure shows that this region of the disk is in equilibrium until approximately 150 years before the outburst. Then the system observes an increase in the aggregate parameter $\alpha_{\rm visc} \cdot \Sigma_{g} \cdot \tau$, and this increase is quite steep (for convenience of comparison, the yellow dashed line shows a function proportional to the cube of the temperature, and the green line shows a function proportional to the temperature to the tenth power, obtained as a result of data approximation). Since the dependence increases at a greater angle than cubic, the development of thermal instability is a likely scenario at this stage. The temperature rises until the threshold of thermal ionization of the medium is exceeded. As a result, an MRI outbreak occurs, during which the rate of viscous heating sharply increases, and then the system relaxes in a quasi-equilibrium state, in which the decline is in good agreement with the cubic law.

If we take a closer look at the components that determine equilibrium, namely, the rate of viscous heating, which can be approximately written as:
\begin{equation}
\Gamma_{\rm visc} = \dfrac{9}{4}\Omega_{\rm K}^{2} \Sigma_{\rm g} \nu_{\rm kin},
\end{equation}
where $\Omega_{\rm K}$ is the angular velocity of the disk, and $\nu_{\rm kin}$ — kinematic viscosity.

In turn, the rate of cooling by radiation is defined in the model as \citep{Dong2016}:
\begin{equation}
\lambda = \dfrac{8 \tau_{\rm P} \sigma T^{4} }{1 + 2\tau_{\rm P} + 1.5\tau_{\rm P} \tau_{\rm R}},
\end{equation}
Where $\tau_{\rm R}$ And $\tau_{\rm P}$ are the optical thicknesses of Rosseland and Planck, respectively, and $\sigma$ – Stefan-Boltzmann constant. From the condition of equilibrium of these quantities, for example, it is possible to determine the critical value of surface density:
\begin{equation}
\Sigma_{\rm g}^{\rm crit} = \dfrac{32}{9} \cdot \dfrac{\mu \sigma \tau_{\rm P} T^{3}}{\alpha_{\rm visc} \gamma R \Omega_{\rm K}} \cdot \dfrac{1}{1 + 2\tau_{\rm P} + 1.5\tau_{\rm P} \tau_{\rm R}},
\end{equation}
where $\gamma$—adiabatic exponent, which in the current study is taken equal to $7 / 5$, which corresponds to a diatomic ideal gas, $\mu=2.33$ is the molecular weight, and ${\rm R}$ is the universal gas constant. If exceeded $\Sigma_{\rm g}^{\rm crit}$ for current values of the viscosity parameter and temperature, the disk should go into a thermal instability mode, provided that its state in the phase space “$T - \Sigma_{\rm g}$” is located in the vicinity of the unstable branch of the S-shaped curve. The results of calculating this critical value are shown in Fig.\ref{fig:critdens}, together with the real azimuthally averaged surface density $\Sigma_{\rm g}$. From the very beginning of the calculation, the surface density is on average an order of magnitude less than the critical value. However, the value of the critical value drops noticeably at a time of about 13 thousand years from the beginning of the calculation, and the real value of the surface density, thus, turns out to be higher than the critical level. At this moment, thermal instability is initiated in the region of the disk under consideration, shortly after which an MRI burst of luminosity develops.

\subsection{Features of Thermal Instability in a Model with a Perturbed Disk}
\begin{figure}
\includegraphics[width = \linewidth]{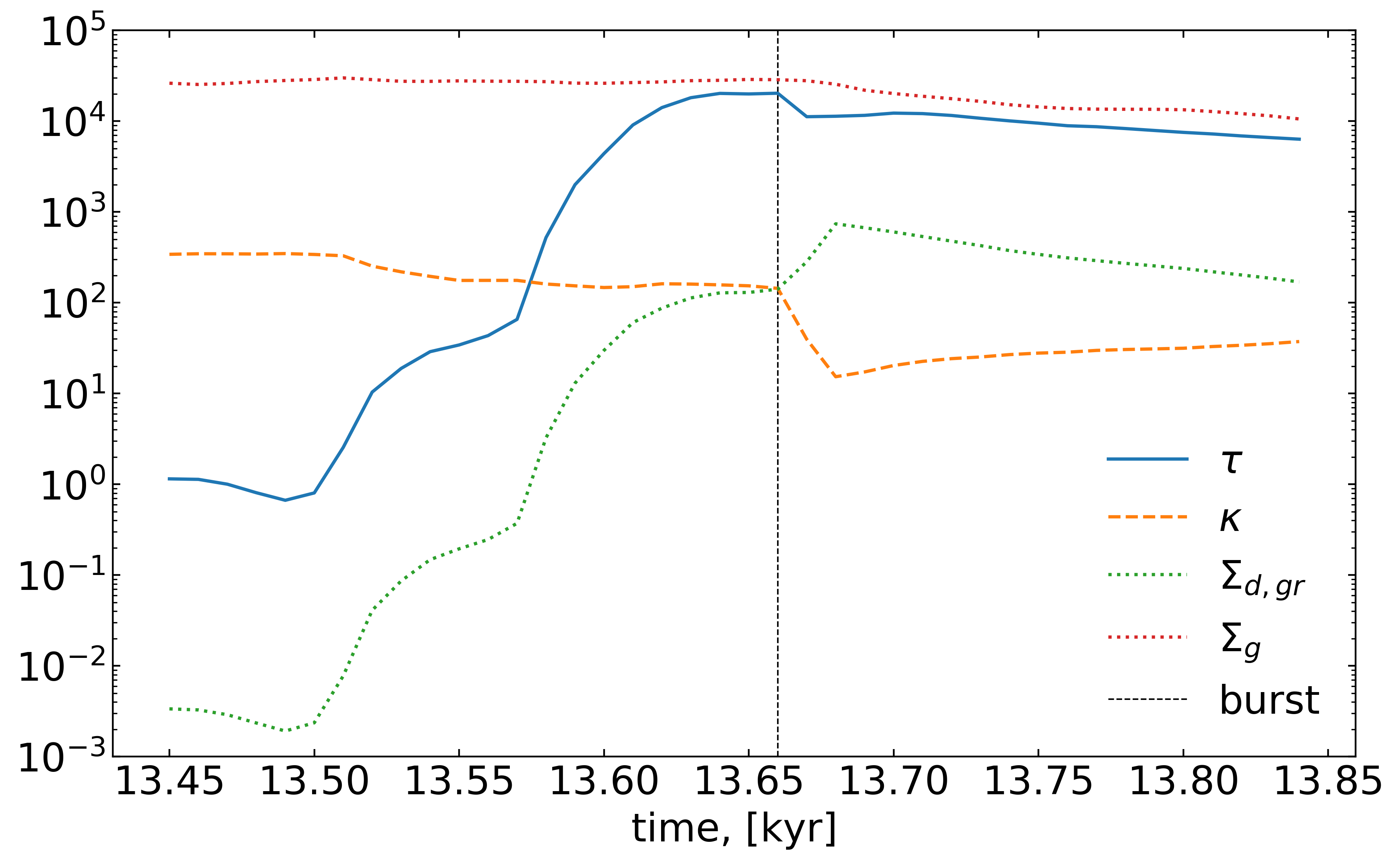}
\caption{Various parameters in a perturbed disk at a distance of 0.75~AU from the center as a function of time. Solid blue line— optical thickness, orange dashed line—opacity in cm$^{2}$/g, green and red dotted lines are the surface densities of the grown dust and gas, respectively, in g/cm$^{2}$. The vertical black dashed line shows the moment of initiation of the MRI outbreak.}
\label{fig:preheat}
\end{figure}

\begin{figure*}
\includegraphics[width = \textwidth]{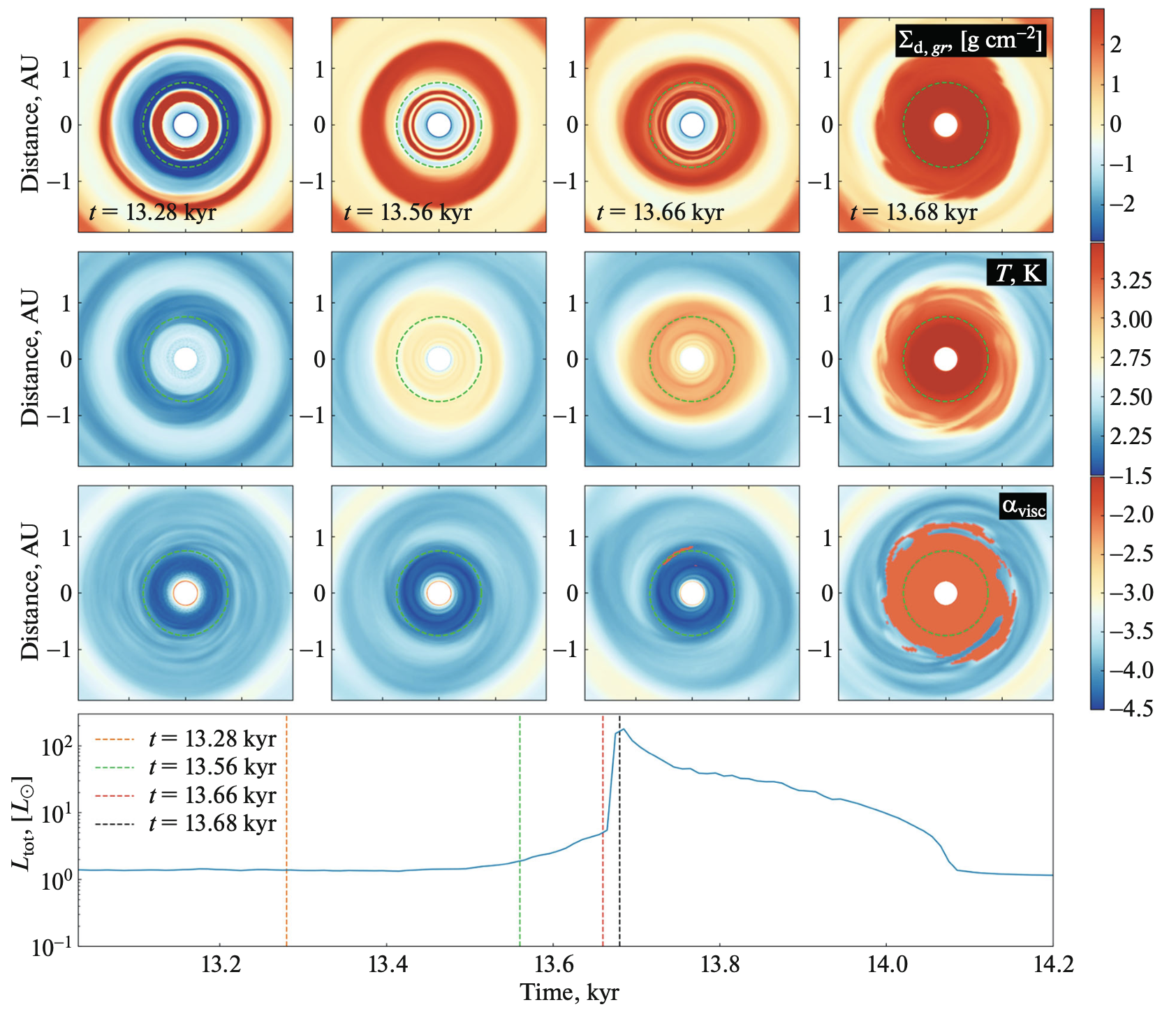}
\caption{Time evolution of distributions of surface density of grown dust (top row), temperature (second row) and viscosity parameter $\alpha_{\rm visc}$ (third row) in the inner 4 a.u. of the disk. The distributions at different moments of the disk evolution are shown, from left to right: t = 13.28, 13.56, 13.66, 13.68 thousand years. The green dashed line indicates the region under consideration, located at a distance of 0.75 AU from the central star. Values on a logarithmic scale correspond to the color bar (right). The bottom panel shows the total luminosity of the central star as a function of time. Vertical dashed lines indicate the time points shown in the bivariate distributions.}
\label{fig:ringfall}
\end{figure*}

The mechanism for the development of thermal instability described above gives an idea of why the MRI and subsequent accretion burst are initiated. However, in the classical sense, thermal instability, described in \cite{BellLin1994, Hartmann1998} and many others, operates in a completely different temperature regime. The characteristic S-shaped shape of the equilibrium curve appears at temperatures above 2–3 thousand K. This feature is a consequence of changes in the optical thickness of the disk, since at such high temperatures hydrogen ionization occurs, which changes the opacity of the medium. However, in our case, the temperatures are an order of magnitude lower. Therefore, the above-mentioned changes in opacity are not the cause of the conditions for the development of thermal instability in our model.

Figure~\ref{fig:preheat} shows the time evolution of various disk parameters before, during, and after the luminosity outburst. It is obvious that by the time the outburst occurs, there is a significant increase in the optical thickness. However, the opacity undergoes only negligible changes. However, it is worth noting that the optical thickness depends not only on the characteristics of the material (gas or dust), but also on the amount of this material itself. Essentially, in our case, the optical thickness is the product of the opacity and the surface density of the dust, since at the temperatures studied, it is the dust that is the main source of opacity and, therefore, the optical thickness. In addition, as can be seen from the figure, the surface density of the gas in the region under consideration also does not undergo significant changes. Thus, it is the increase in the surface density of dust that causes the increase in optical thickness and, consequently, the creation of conditions for the development of thermal instability.

The increase in the surface density of dust in the region under consideration is illustrated in Fig.~\ref{fig:ringfall}. Study area at a distance $r=0.75$~AU. in which thermal instability first develops is shown by the green dashed line. Initially, there is an extremely small amount of grown dust in this area, because it is rather a gap between two ring structures, internal and external. These ring structures contain quite a lot of dust and are generally quite stable. The temperature in them is low for the development of classical thermal (due to hydrogen ionization) or magnetorotational instabilities, due to effective cooling due to low optical thicknesses, and the viscosity at this radius undergoes a transition from a low value corresponding to the dead zone to a higher one. The disturbance caused by the close passage of an external object leads to an imbalance in the system. The rings located outside the region under consideration lose their initially regular shape (see Fig.~\ref{fig:initial}), transform into elliptical structures and begin to precess, which is clearly noticeable in the second and third columns of the panels. Thus, the ring, located at a distance of about 2 AU, gradually drifts towards the center towards the more stable inner ring. This process leads to a local increase in optical thickness and the subsequent development of thermal instability, which is correspondingly expressed in an increase in temperature in the region under study. Ultimately, upon reaching the critical temperature value at the time $t=13.66$ thousand years, MRI is locally activated, which can be seen in the panel in the third row of the third column. Soon after this, the MRI covers the entire inner disk, and a burst of luminosity develops, caused by a significant increase in the accretion rate, and the ring system is completely destroyed. After the decay of the MRI and the outbreak of accretion, a rotating vortex structure appears in the system, which over time degenerates into a dust ring localized in the region where the outer ring is present in the pre-outburst state.

\section{DISCUSSION OF RESULTS AND LIMITATIONS OF THE MODEL}
First of all, it is worth noting that for the implementation of the presented cascade mechanism for triggering an outburst, the presence of dust rings and gaps in the inner regions of the disk is critical. The method of forming such structures and their type may be different. Rings in the vicinity of ice lines can form due to different sizes of dust inside and outside the line and, accordingly, different velocities of radial drift \citep{Zhang2015, Pinilla2017}. The appearance of ring and gap structures in the disk can also be caused by a planet \citep{2015Kley, Dong2015}. In the case we considered, rings form in the dead zones of the disk, which are natural dust traps \citep{Dzyurkevich2010, Flock2015, Kadam2022, Vorobyov2023}. In the simulated “layered” disk, the different efficiency of matter transfer by viscous force moments—more efficient outside and ineffective inside the dead zone— leads to the development of a situation reminiscent of a narrow bottleneck. The inability to transfer the entire volume of matter coming into the dead zone from the outer disk leads to its accumulation and an increase in pressure, up to the formation of a local peak. Dust drifting in the direction of the pressure gradient inevitably accumulates in such peaks, which is expressed in the appearance of rings whose position coincides with the peaks, and gaps in the areas between the peaks, if there are several of them and they are localized close enough. Separately, it can be noted that dead zones can develop not only as a result of different efficiency of MRI, as implied in the used model of a layered disk, but also as a consequence of weak turbulent viscosity and variable efficiency of gravitational transfer in the disk \citep{Vorobyov2023}.

Despite the abundance of opportunities for the formation of dust rings in disks, the rings presented in this paper are quite unlikely to be observed. On the one hand, interferometric observations make it possible to obtain information about the distribution of relatively large dust, which settles towards the middle plane of the disk and can effectively accumulate into rings. On the other hand, the capabilities of modern (sub)millimeter interferometers do not allow spatial resolution of such narrow rings in such a close neighborhood of the star, as presented in the current work. Scattered light observations might allow such structures to be resolved, but during the interstitial phase the shell contains sufficient amounts of fine, submicron dust to naturally screen radiation emanating from the midplane in which the rings are expected to be localized. In addition, submicron dust itself does not drift to pressure maxima, but is tightly bound to the gas. As a consequence, the spatial segregation of submicron dust is not as pronounced as that of millimeter dust, which further reduces the chances of successful detection of dust rings, which play a key role in the presented mechanism of outburst development.

The shown scenario of cascade initialization of an outburst is of interest, among other things, because it has features similar to an outburst in the FU Orionis system. First of all, the amplitudes of the outbursts are consistent—the luminosity increases by 2 orders of magnitude in a short period of time. As in the case of FU Orionis, the system is a binary system, consisting of objects of different masses. Just like in FU Orionis, the outburst occurs on a less massive star \citep{BeckAspin2012}. Finally, apparently, the flyby, like in FU Orionis, is non-penetrating. Despite these similarities, the purpose of the presented work is not to directly explain the outburst scenario of a specific object like FU Orionis, but to explore possible mechanisms by which outbursts can occur in systems with relatively large periastrons and/or delays. Unfortunately, it is not possible to reliably state that the cascaded initialization shown took place in the FU Orionis. In addition, the parameters of the systems still differ: the periastron in the simulated binary system is twice as large as even the current distance between objects in FU Orionis \citep{Perez2020}, and the distance at the moment of the outburst is much greater. In addition, the size of the disk around the central star (>100 AU) exceeds those in FU Orionis \citep{Perez2020}, and although the masses of the outburst stars are similar ($0.28 \  M_{\odot}$ in the model system and $\sim 0.3 \ M_{\odot}$ in FU Orionis), the masses of stars in the quiet phase differ by approximately 2 times: $0.6 \ M_{\odot}$ and $1.2 \ M_{\odot}$ respectively \citep{BeckAspin2012}. However, the main task of finding opportunities to implement a flash during non-penetrating flights, as well as with a delay, has been completed. 

Despite the subtlety of the conditions necessary for the implementation of the shown cascade development of an outburst: the presence of dust rings and gaps in the inner disk, as well as their presence in a state of marginal stability, in which the gravitational disturbance from the passage of an external object will be able to unbalance the system, the proposed mechanism can expand the number of binary systems prone to outburst development due to passage. The fact is that approaches with large periastrons are more likely events. In \cite{ForganRice2010}, an estimate based on calculations \cite{ClarkePringle1991} is given, which makes it possible to calculate the frequency of encounter events between a star-disk system and diskless companions in a young star cluster within a certain radius $R_{\rm enc}$:
\begin{equation} 
\label{eq:G1}
\Gamma_{\rm hit} = \Gamma_{0} \left(1 + \dfrac{V_{*}^{2} R_{\rm enc}}{G M_{\ast}} \right),
\end{equation}
their velocity distribution is assumed to be Gaussian with dispersion $V_{\ast}$, $\Gamma_{0}$, in turn, is defined as:
\begin{equation} 
\label{eq:G0}
\Gamma_{0} = \dfrac{4 \sqrt{\pi} n_{0} G M_{\ast} R_{\rm enc}}{V_{\ast}}.
\end{equation} 
where $n_{0}$—concentration of stars in the cluster. Let’s take typical values for a young star cluster $n_{0} = 100$~ PC$^{-3}$ And $V_{\ast} = 1$~km/s \citep{ClarkePringle1991}. The mass of the studied disk around the central star in the current work is $0.22 \ M_{\odot}$, and as $R_{\rm enc}$ let’s choose two characteristic values to compare the probability of passage with different radii. Let’s take one characteristic value equal to 20~AU, which approximately corresponds to a close penetrating passage, as a result of which an outburst can occur \citep{Cuello2023}. Taking into account that the average age of the protostellar disk is approximately 1 million years, and by substituting the values in equations~\eqref{eq:G0},~\eqref{eq:G1}, we can calculate that a collision with such parameters can occur in approximately one star-disk system from 11000. If, as a $R_{\rm enc}$ if we take the periastron of the system modeled in this work with cascade initiation of an outburst equal to 500 AU, the result will be more likely—one system out of 157. Thus, the probability of collisions such as the one studied in the current work is approximately 70 times higher than the typical penetrating case in compact disk systems. In addition, since the proportion of binary and multiple systems increases with the mass of the star, one can expect that the relevance of the presented outburst mechanism will increase with the mass of the central star. In addition, it can be mentioned that the shown mechanism of cascade initiation of an outburst can take place not only in binary systems, but also, for example, in the case of the fall of clumps of matter from the remnants of the parent cloud. This possibility was also noted in \cite{Demidova2023}. 

Finally, we will pay attention to the process of outburst development from the point of view of localization of outburst processes. As described above, the outburst occurs due to the development of thermal and magnetorotational instabilities caused by the imbalance of the system due to gravitational disturbance by an external object. Thermal instability heats the system sufficiently to initiate MRI and subsequent significant increases in the accretion rate and luminosity of the central star. However, there are different results showing the spatial direction of the propagation of instabilities. For example, in \cite{Bourdarot2023}, it is shown that an outburst develops due to the activation of the MRI, first at the outer boundary of the dead zone due to the effective influx of matter through the transfer of gravitational force moments from gravitationally unstable regions of the disk inward. Then, the MRI gradually develops closer to the star, where local development of thermal instability is already possible, accompanied by a significant increase in the accretion rate and a burst of luminosity. On the other hand, in the simulation \cite{Kadam2021}, on the contrary, the MRI propagates from the inside to the outside due to the gradual achievement of the threshold of thermal ionization by the substance in the direction from hotter to less heated regions. In the study \citep{Bae2014}, the direction of propagation depends on the value of the background viscosity in the dead zone. In the current work, it was not possible to examine in detail the mechanics of the development of instability due to the insufficient frequency of temporary step, and the solution to this problem can become the basis for subsequent research. However, as shown in Fig.~\ref{fig:pre-burst}, at the initial stage, MCI develops to a greater extent at the outer boundary of the dead zone. At the very next output of time data, the MRI already covers the entire dead zone and some area outside. A detailed description of the process requires extremely frequent data output, since in these areas the characteristic viscous and dynamic times are small.

\section{CONCLUSIONS}
\label{Sect:conclude}
Numerical studies of FU Orionis-type outbursts help answer questions about the origin and general characteristics of these phenomena, but in reality the causes of outbursts can be different. In particular, this work investigated a scenario in which the triggering mechanism for a luminosity outburst is a close passage of an external disturbing stellar object, but the immediate mechanism for the development and maintenance of the outburst is thermal and magnetorotational instabilities in the inner regions of the protoplanetary disk. The main results of the study can be highlighted as follows:
\begin{itemize}
    \item The primary disturbance caused by the gravitational influence of an external object upon approach may not provoke significant changes in the luminosity of the central object on a scale comparable to fuor outbursts. In the system studied, the shortest distance between objects was about 500 AU. The disturbance took more than 800 years to reach the protostellar accretion region of the central object. The luminosity of the central object increased by approximately 2 times, due to disturbances caused by the primary gravitational influence from the flyby.
    \item The fly-by of an external object may not directly, but indirectly, cause a much more powerful outburst. More than 5 thousand years after the flyby, conditions for the development of thermal and MRI were realized in the disk under study, which led to an increase in the accretion rate and a corresponding increase in the luminosity of the central star by more than 100 times over a short time scale of the phenomenon (less than 10 years).
    \item For the first time, numerical simulations have shown that a gravitational perturbation resulting from the fly-by of a star can lead to an imbalance in the disk of the central star and trigger a cascade of effects, ultimately leading to an FU Orionis-type outburst. The chain of events is as follows: the initial system of dust rings was deformed, precession developed, as a result of which one of the rings was destroyed, followed by a local increase in optical depth and the resulting development of thermal instability. Thermal instability, leading to an increase in temperature to values above critical values, in turn, triggered MRI, which significantly increased the accretion rate and, accordingly, the luminosity of the star.
\end{itemize}

\section{ACKNOWLEDGMENTS}
The work was supported by the Ministry of Science and Higher Education of the Russian Federation, State assignment in the field of scientific activity no. GZ0110/23-10-IF (A.M.S., Sections 1, 3, 4) and the Russian Science Foundation, project No. 23-12- 00258 (E.I.V., Sections 2, 5).

%%\end{enumerate}

\begin{acknowledgements}
%This work was supported by the
%Ministry of Science and Higher Education of the Russian
%Federation (State assignment in the field of scientific activity 2023, GZ0110/23-10-IF).
%Simulations were performed on the Vienna Scientific Cluster (VSC) \footnote{\href{https://vsc.ac.at/}{https://vsc.ac.at/}}.
\end{acknowledgements}

\bibliographystyle{aa}
\bibliography{refs}

\begin{thebibliography}{65}
\expandafter\ifx\csname natexlab\endcsname\relax\def\natexlab#1{#1}\fi

\bibitem[{{Armitage} {et~al.}(2001){Armitage}, {Livio}, \&
  {Pringle}}]{Armitage2001}
{Armitage}, P.~J., {Livio}, M., \& {Pringle}, J.~E. 2001, Mon. Not. Roy.
  Astron. Soc., 324, 705

\bibitem[{{Audard} {et~al.}(2014){Audard}, {{\'A}brah{\'a}m}, {Dunham},
  {Green}, {Grosso}, {Hamaguchi}, {Kastner}, {K{\'o}sp{\'a}l}, {Lodato},
  {Romanova}, {Skinner}, {Vorobyov}, \& {Zhu}}]{Audard2014}
{Audard}, M., {{\'A}brah{\'a}m}, P., {Dunham}, M.~M., {et~al.} 2014, in
  Protostars and Planets VI, ed. H.~{Beuther}, R.~S. {Klessen}, C.~P.
  {Dullemond}, \& T.~{Henning}, 387

\bibitem[{{Bae} {et~al.}(2014){Bae}, {Hartmann}, {Zhu}, \& {Nelson}}]{Bae2014}
{Bae}, J., {Hartmann}, L., {Zhu}, Z., \& {Nelson}, R.~P. 2014, \apj, 795, 61

\bibitem[{{Banzatti} {et~al.}(2015){Banzatti}, {Pinilla}, {Ricci},
  {Pontoppidan}, {Birnstiel}, \& {Ciesla}}]{Banzatti2015}
{Banzatti}, A., {Pinilla}, P., {Ricci}, L., {et~al.} 2015, Astrophys. J. Lett.,
  815, L15

\bibitem[{{Beck} \& {Aspin}(2012)}]{BeckAspin2012}
{Beck}, T.~L. \& {Aspin}, C. 2012, The Astronomical Journal, 143, 55

\bibitem[{{Bell} \& {Lin}(1994)}]{BellLin1994}
{Bell}, K.~R. \& {Lin}, D.~N.~C. 1994, \apj, 427, 987

\bibitem[{{Bonnell} \& {Bastien}(1992)}]{BonnellBastien1992}
{Bonnell}, I. \& {Bastien}, P. 1992, \apj Letters, 401, L31

\bibitem[{{Borchert} {et~al.}(2022{\natexlab{a}}){Borchert}, {Price}, {Pinte},
  \& {Cuello}}]{Borchert2022}
{Borchert}, E. M.~A., {Price}, D.~J., {Pinte}, C., \& {Cuello}, N.
  2022{\natexlab{a}}, Monthly Notices Roy. Astron. Soc., 510, L37

\bibitem[{{Borchert} {et~al.}(2022{\natexlab{b}}){Borchert}, {Price}, {Pinte},
  \& {Cuello}}]{Borchert2022b}
{Borchert}, E. M.~A., {Price}, D.~J., {Pinte}, C., \& {Cuello}, N.
  2022{\natexlab{b}}, Mon. Not. Roy. Astron. Soc., 517, 4436

\bibitem[{{Bourdarot} {et~al.}(2023){Bourdarot}, {Berger}, {Lesur}, {Perraut},
  {Malbet}, {Millan-Gabet}, {Le Bouquin}, {Garcia-Lopez}, {Monnier}, {Labdon},
  {Kraus}, {Labadie}, \& {Aarnio}}]{Bourdarot2023}
{Bourdarot}, G., {Berger}, J.-P., {Lesur}, G., {et~al.} 2023, arXiv e-prints,
  arXiv:2304.13414

\bibitem[{{Clarke} \& {Pringle}(1991)}]{ClarkePringle1991}
{Clarke}, C.~J. \& {Pringle}, J.~E. 1991, Mon. Not. Roy. Astron. Soc., 249, 584

\bibitem[{{Connelley} \& {Reipurth}(2018)}]{Connelley2018}
{Connelley}, M.~S. \& {Reipurth}, B. 2018, \apj, 861, 145

\bibitem[{{Cuello} {et~al.}(2023){Cuello}, {M{\'e}nard}, \&
  {Price}}]{Cuello2023}
{Cuello}, N., {M{\'e}nard}, F., \& {Price}, D.~J. 2023, European Physical
  Journal Plus, 138, 11

\bibitem[{{D'Alessio}(1996)}]{DAlessio1996}
{D'Alessio}, P. 1996, PhD thesis, UNAM, Institute of Astronomy

\bibitem[{{Demidova} \& {Grinin}(2023)}]{Demidova2023}
{Demidova}, T.~V. \& {Grinin}, V.~P. 2023, \apj, 953, 38

\bibitem[{{Dohnanyi}(1969)}]{Dohnanyi1969}
{Dohnanyi}, J.~S. 1969, Journal of Geophysical Research, 74, 2531

\bibitem[{{Dong} {et~al.}(2022){Dong}, {Liu}, {Cuello}, {Pinte},
  {{\'A}brah{\'a}m}, {Vorobyov}, {Hashimoto}, {K{\'o}sp{\'a}l}, {Chiang},
  {Takami}, {Chen}, {Dunham}, {Fukagawa}, {Green}, {Hasegawa}, {Henning},
  {Pavlyuchenkov}, {Pyo}, \& {Tamura}}]{Dong2022}
{Dong}, R., {Liu}, H.~B., {Cuello}, N., {et~al.} 2022, Nature Astronomy, 6, 331

\bibitem[{{Dong} {et~al.}(2016){Dong}, {Vorobyov}, {Pavlyuchenkov}, {Chiang},
  \& {Liu}}]{Dong2016}
{Dong}, R., {Vorobyov}, E., {Pavlyuchenkov}, Y., {Chiang}, E., \& {Liu}, H.~B.
  2016, \apj, 823, 141

\bibitem[{{Dong} {et~al.}(2015){Dong}, {Zhu}, \& {Whitney}}]{Dong2015}
{Dong}, R., {Zhu}, Z., \& {Whitney}, B. 2015, \apj, 809, 93

\bibitem[{{Dullemond} {et~al.}(2019){Dullemond}, {K{\"u}ffmeier}, {Goicovic},
  {Fukagawa}, {Oehl}, \& {Kramer}}]{Dullemond2019}
{Dullemond}, C.~P., {K{\"u}ffmeier}, M., {Goicovic}, F., {et~al.} 2019, Astron.
  Astrophys., 628, A20

\bibitem[{{Dzyurkevich} {et~al.}(2010){Dzyurkevich}, {Flock}, {Turner},
  {Klahr}, \& {Henning}}]{Dzyurkevich2010}
{Dzyurkevich}, N., {Flock}, M., {Turner}, N.~J., {Klahr}, H., \& {Henning}, T.
  2010, Astron. and Astrophys., 515, A70

\bibitem[{{Flock} {et~al.}(2015){Flock}, {Ruge}, {Dzyurkevich}, {Henning},
  {Klahr}, \& {Wolf}}]{Flock2015}
{Flock}, M., {Ruge}, J.~P., {Dzyurkevich}, N., {et~al.} 2015, Astron. and
  Astrophys., 574, A68

\bibitem[{{Forgan} \& {Rice}(2010)}]{ForganRice2010}
{Forgan}, D. \& {Rice}, K. 2010, Mon. Not. Roy. Astron. Soc., 402, 1349

\bibitem[{{Gammie}(1996)}]{Gammie1996}
{Gammie}, C.~F. 1996, \apj, 457, 355

\bibitem[{{Hartmann}(1998)}]{Hartmann1998}
{Hartmann}, L. 1998, {Accretion Processes in Star Formation}

\bibitem[{{Henderson}(1976)}]{Henderson1976}
{Henderson}, C.~B. 1976, AIAA Journal, 14, 707

\bibitem[{{Kadam} {et~al.}(2022){Kadam}, {Vorobyov}, \& {Basu}}]{Kadam2022}
{Kadam}, K., {Vorobyov}, E., \& {Basu}, S. 2022, Monthly Notices Roy. Astron.
  Soc., 516, 4448

\bibitem[{{Kadam} {et~al.}(2021){Kadam}, {Vorobyov}, \&
  {K{\'o}sp{\'a}l}}]{Kadam2021}
{Kadam}, K., {Vorobyov}, E., \& {K{\'o}sp{\'a}l}, {\'A}. 2021, \apj, 909, 31

\bibitem[{{Kadam} {et~al.}(2019){Kadam}, {Vorobyov}, {Reg{\'a}ly},
  {K{\'o}sp{\'a}l}, \& {{\'A}brah{\'a}m}}]{Kadam2019}
{Kadam}, K., {Vorobyov}, E., {Reg{\'a}ly}, Z., {K{\'o}sp{\'a}l}, {\'A}., \&
  {{\'A}brah{\'a}m}, P. 2019, \apj, 882, 96

\bibitem[{{Kadam} {et~al.}(2020){Kadam}, {Vorobyov}, {Reg{\'a}ly},
  {K{\'o}sp{\'a}l}, \& {{\'A}brah{\'a}m}}]{Kadam2020}
{Kadam}, K., {Vorobyov}, E., {Reg{\'a}ly}, Z., {K{\'o}sp{\'a}l}, {\'A}., \&
  {{\'A}brah{\'a}m}, P. 2020, \apj, 895, 41

\bibitem[{{Kawazoe} \& {Mineshige}(1993)}]{Kawazoe1993}
{Kawazoe}, E. \& {Mineshige}, S. 1993, Pub. of the Astron. Society of Japan,
  45, 715

\bibitem[{{Kenyon}(1999)}]{Kenyon1999}
{Kenyon}, S.~J. 1999, in NATO Advanced Study Institute (ASI) Series C, Vol.
  540, The Origin of Stars and Planetary Systems, ed. C.~J. {Lada} \& N.~D.
  {Kylafis}, 613

\bibitem[{{Kuffmeier} {et~al.}(2018){Kuffmeier}, {Frimann}, {Jensen}, \&
  {Haugb{\o}lle}}]{Kuffmeier2018}
{Kuffmeier}, M., {Frimann}, S., {Jensen}, S.~S., \& {Haugb{\o}lle}, T. 2018,
  Mon. Not. Roy. Astron. Soc., 475, 2642

\bibitem[{{Labdon} {et~al.}(2021){Labdon}, {Kraus}, {Davies}, {Kreplin},
  {Monnier}, {Le Bouquin}, {Anugu}, {ten Brummelaar}, {Setterholm}, {Gardner},
  {Ennis}, {Lanthermann}, {Schaefer}, \& {Laws}}]{Labdon2021}
{Labdon}, A., {Kraus}, S., {Davies}, C.~L., {et~al.} 2021, Astron. Astrophys.,
  646, A102

\bibitem[{{Liu} {et~al.}(2017){Liu}, {Vorobyov}, {Dong}, {Dunham}, {Takami},
  {Galv{\'a}n-Madrid}, {Hashimoto}, {K{\'o}sp{\'a}l}, {Henning}, {Tamura},
  {Rodr{\'\i}guez}, {Hirano}, {Hasegawa}, {Fukagawa}, {Carrasco-Gonzalez}, \&
  {Tazzari}}]{Liu2017}
{Liu}, H.~B., {Vorobyov}, E.~I., {Dong}, R., {et~al.} 2017, Astron. Astrophys.,
  602, A19

\bibitem[{{Lykou} {et~al.}(2022){Lykou}, {{\'A}brah{\'a}m}, {Chen}, {Varga},
  {K{\'o}sp{\'a}l}, {Matter}, {Siwak}, {Szab{\'o}}, {Zhu}, {Liu}, {Lopez},
  {Allouche}, {Augereau}, {Berio}, {Cruzal{\`e}bes}, {Dominik}, {Henning},
  {Hofmann}, {Hogerheijde}, {Jaffe}, {Kokoulina}, {Lagarde}, {Meilland},
  {Millour}, {Pantin}, {Petrov}, {Robbe-Dubois}, {Schertl}, {Scheuck}, {van
  Boekel}, {Waters}, {Weigelt}, \& {Wolf}}]{Lykou2022}
{Lykou}, F., {{\'A}brah{\'a}m}, P., {Chen}, L., {et~al.} 2022, Astron.
  Astrophys., 663, A86

\bibitem[{{Magakian} {et~al.}(2022){Magakian}, {Movsessian}, \&
  {Andreasyan}}]{Magakian2022}
{Magakian}, T., {Movsessian}, T., \& {Andreasyan}, H. 2022, Acta Astrophysica
  Taurica, 3, 4

\bibitem[{{Maksimova} {et~al.}(2020){Maksimova}, {Pavlyuchenkov}, \&
  {Tutukov}}]{Maksimova2020}
{Maksimova}, L.~A., {Pavlyuchenkov}, Y.~N., \& {Tutukov}, A.~V. 2020, Astronomy
  Reports, 64, 815

\bibitem[{{Mercer} \& {Stamatellos}(2017)}]{Mercer2017}
{Mercer}, A. \& {Stamatellos}, D. 2017, Monthly Notices Roy. Astron. Soc., 465,
  2

\bibitem[{{Molyarova} {et~al.}(2018){Molyarova}, {Akimkin}, {Semenov},
  {{\'A}brah{\'a}m}, {Henning}, {K{\'o}sp{\'a}l}, {Vorobyov}, \&
  {Wiebe}}]{Molyarova2018}
{Molyarova}, T., {Akimkin}, V., {Semenov}, D., {et~al.} 2018, \apj, 866, 46

\bibitem[{{Molyarova} {et~al.}(2021){Molyarova}, {Vorobyov}, {Akimkin},
  {Skliarevskii}, {Wiebe}, \& {G{\"u}del}}]{Molyarova2021}
{Molyarova}, T., {Vorobyov}, E.~I., {Akimkin}, V., {et~al.} 2021, \apj, 910,
  153

\bibitem[{{Nayakshin} \& {Lodato}(2012)}]{Nayakshin2012}
{Nayakshin}, S. \& {Lodato}, G. 2012, Mon. Not. Roy. Astron. Soc., 426, 70

\bibitem[{{P{\'e}rez} {et~al.}(2020){P{\'e}rez}, {Hales}, {Liu}, {Zhu},
  {Casassus}, {Williams}, {Zurlo}, {Cuello}, {Cieza}, \&
  {Principe}}]{Perez2020}
{P{\'e}rez}, S., {Hales}, A., {Liu}, H.~B., {et~al.} 2020, \apj, 889, 59

\bibitem[{{Picogna} \& {Kley}(2015)}]{2015Kley}
{Picogna}, G. \& {Kley}, W. 2015, Astron. and Astrophys., 584, A110

\bibitem[{{Pinilla} {et~al.}(2017){Pinilla}, {Pohl}, {Stammler}, \&
  {Birnstiel}}]{Pinilla2017}
{Pinilla}, P., {Pohl}, A., {Stammler}, S.~M., \& {Birnstiel}, T. 2017, \apj,
  845, 68

\bibitem[{{Rab} {et~al.}(2017){Rab}, {Elbakyan}, {Vorobyov}, {G{\"u}del},
  {Dionatos}, {Audard}, {Kamp}, {Thi}, {Woitke}, \& {Postel}}]{Rab2017}
{Rab}, C., {Elbakyan}, V., {Vorobyov}, E., {et~al.} 2017, Astron. Astrophys.,
  604, A15

\bibitem[{{Reg{\'a}ly} \& {Vorobyov}(2017)}]{2017RegalyVorobyov}
{Reg{\'a}ly}, Z. \& {Vorobyov}, E. 2017, Monthly Notices Roy. Astron. Soc.,
  471, 2204

\bibitem[{{Schoonenberg} \& {Ormel}(2017)}]{Schoonenberg2017}
{Schoonenberg}, D. \& {Ormel}, C.~W. 2017, Astron. Astrophys., 602, A21

\bibitem[{{Shakura} \& {Sunyaev}(1973)}]{1973ShakuraSunyaev}
{Shakura}, N.~I. \& {Sunyaev}, R.~A. 1973, Astron. and Astrophys.Astron. and
  Astrophys., 24, 337

\bibitem[{{Skliarevskii} {et~al.}(2021){Skliarevskii}, {Pavlyuchenkov}, \&
  {Vorobyov}}]{Skliarevskii2021}
{Skliarevskii}, A.~M., {Pavlyuchenkov}, Y.~N., \& {Vorobyov}, E.~I. 2021,
  Astronomy Reports, 65, 170

\bibitem[{{Stoyanovskaya} {et~al.}(2020){Stoyanovskaya}, {Okladnikov},
  {Vorobyov}, {Pavlyuchenkov}, \& {Akimkin}}]{Stoyanovskaya2020}
{Stoyanovskaya}, O.~P., {Okladnikov}, F.~A., {Vorobyov}, E.~I.,
  {Pavlyuchenkov}, Y.~N., \& {Akimkin}, V.~V. 2020, Astronomy Reports, 64, 107

\bibitem[{{Visser} {et~al.}(2015){Visser}, {Bergin}, \&
  {J{\o}rgensen}}]{Visser2015}
{Visser}, R., {Bergin}, E.~A., \& {J{\o}rgensen}, J.~K. 2015, Astron.
  Astrophys., 577, A102

\bibitem[{{Vorobyov} {et~al.}(2018){Vorobyov}, {Akimkin}, {Stoyanovskaya},
  {Pavlyuchenkov}, \& {Liu}}]{2018VorobyovAkimkin}
{Vorobyov}, E.~I., {Akimkin}, V., {Stoyanovskaya}, O., {Pavlyuchenkov}, Y., \&
  {Liu}, H.~B. 2018, Astron. and Astrophys., 614, A98

\bibitem[{{Vorobyov} {et~al.}(2013){Vorobyov}, {Baraffe}, {Harries}, \&
  {Chabrier}}]{VorobyovBaraffe2013}
{Vorobyov}, E.~I., {Baraffe}, I., {Harries}, T., \& {Chabrier}, G. 2013,
  Astron. Astrophys., 557, A35

\bibitem[{{Vorobyov} \& {Basu}(2010)}]{2010VorobyovBasu}
{Vorobyov}, E.~I. \& {Basu}, S. 2010, ApJ, 719, 1896

\bibitem[{{Vorobyov} \& {Basu}(2015)}]{VorobyovBasu2015}
{Vorobyov}, E.~I. \& {Basu}, S. 2015, \apj, 805, 115

\bibitem[{{Vorobyov} {et~al.}(2023){Vorobyov}, {Elbakyan}, {Johansen},
  {Lambrechts}, {Skliarevskii}, \& {Stoyanovskaya}}]{Vorobyov2023}
{Vorobyov}, E.~I., {Elbakyan}, V.~G., {Johansen}, A., {et~al.} 2023, Astron.
  Astrophys., 670, A81

\bibitem[{{Vorobyov} {et~al.}(2021){Vorobyov}, {Elbakyan}, {Liu}, \&
  {Takami}}]{Vorobyov2021}
{Vorobyov}, E.~I., {Elbakyan}, V.~G., {Liu}, H.~B., \& {Takami}, M. 2021,
  Astron. Astrophys., 647, A44

\bibitem[{{Vorobyov} {et~al.}(2020{\natexlab{a}}){Vorobyov}, {Elbakyan},
  {Takami}, \& {Liu}}]{VorobyovElbakyan2020}
{Vorobyov}, E.~I., {Elbakyan}, V.~G., {Takami}, M., \& {Liu}, H.~B.
  2020{\natexlab{a}}, Astron. and Astrophys., 643, A13

\bibitem[{{Vorobyov} {et~al.}(2020{\natexlab{b}}){Vorobyov}, {Khaibrakhmanov},
  {Basu}, \& {Audard}}]{VorobyovKhaibrakhmanov2020}
{Vorobyov}, E.~I., {Khaibrakhmanov}, S., {Basu}, S., \& {Audard}, M.
  2020{\natexlab{b}}, Astron. and Astrophys., 644, A74

\bibitem[{{Vorobyov} {et~al.}(2014){Vorobyov}, {Pavlyuchenkov}, \&
  {Trinkl}}]{Vorobyov2014}
{Vorobyov}, E.~I., {Pavlyuchenkov}, Y.~N., \& {Trinkl}, P. 2014, Astronomy
  Reports, 58, 522

\bibitem[{{Vorobyov} {et~al.}(2022){Vorobyov}, {Skliarevskii}, {Molyarova},
  {Akimkin}, {Pavlyuchenkov}, {K{\'o}sp{\'a}l}, {Liu}, {Takami}, \&
  {Topchieva}}]{Vorobyov2022}
{Vorobyov}, E.~I., {Skliarevskii}, A.~M., {Molyarova}, T., {et~al.} 2022,
  Astron. and Astrophys., 658, A191

\bibitem[{{Vorobyov} {et~al.}(2017){Vorobyov}, {Steinrueck}, {Elbakyan}, \&
  {Guedel}}]{Vorobyov2017}
{Vorobyov}, E.~I., {Steinrueck}, M.~E., {Elbakyan}, V., \& {Guedel}, M. 2017,
  Astron. Astrophys., 608, A107

\bibitem[{{Wiebe} {et~al.}(2019){Wiebe}, {Molyarova}, {Akimkin}, {Vorobyov}, \&
  {Semenov}}]{Wiebe2019}
{Wiebe}, D.~S., {Molyarova}, T.~S., {Akimkin}, V.~V., {Vorobyov}, E.~I., \&
  {Semenov}, D.~A. 2019, Mon. Not. Roy. Astron. Soc., 485, 1843

\bibitem[{{Zhang} {et~al.}(2015){Zhang}, {Blake}, \& {Bergin}}]{Zhang2015}
{Zhang}, K., {Blake}, G.~A., \& {Bergin}, E.~A. 2015, Astrophys. J. Lett., 806,
  L7

\end{thebibliography}

\end{document}